\documentclass[sigconf,]{acmart}
\AtBeginDocument{%
  }

\setcopyright{acmlicensed}
\copyrightyear{2026}
\acmYear{2026}
\acmDOI{XXXXXXX.XXXXXXX}
\acmISBN{978-1-4503-XXXX-X/2018/06}

\usepackage{soul}
\usepackage{url}
\usepackage[utf8]{inputenc}
\usepackage{caption}
\usepackage{graphicx}
\usepackage{amsmath}
\usepackage{amsthm}
\usepackage{booktabs}
\usepackage{algorithm}
\usepackage{algorithmic}

\usepackage{tikz}
\newcommand*\circled[1]{\tikz[baseline=(char.base)]{
            \node[shape=circle,draw,line width=0.2pt,inner sep=0.6pt] (char) {\scriptsize #1};}}
\usepackage{mdframed}
\usepackage{amsfonts}
\usepackage{makecell}
\usepackage[table]{xcolor}
\definecolor{BestRed}{HTML}{C00000}
\definecolor{SecondPurple}{HTML}{7A1FA2}

\newcommand{\best}[1]{\textbf{\textcolor{BestRed}{#1}}}
\newcommand{\second}[1]{\textcolor{SecondPurple}{#1}}

\usepackage{subcaption}
\usepackage{multirow}
\usepackage{enumitem}
\usepackage[table]{xcolor}

\renewcommand\footnotetextcopyrightpermission[1]{}
\renewcommand{\shortauthors}{Miao et al.}
\begin{document}

\title{UniDetect: LLM-Driven Universal Fraud Detection across Heterogeneous Blockchains}

\author{Shuyi Miao}
\affiliation{%
  \institution{Beihang University}
  \city{Beijing}
  \country{China}
}

\author{Wangjie Qiu}
\authornote{Corresponding authors. Contact email: shuyimiao@buaa.edu.cn}
\affiliation{%
  \institution{Beihang University}
  \city{Beijing}
  \country{China}
}

\author{Shengda Zhuo}
\affiliation{%
  \institution{Jinan University}
  \city{Guangzhou}
  \country{China}
}

\author{Fei Shen}
\authornotemark[1]
\affiliation{%
  \institution{National University of Singapore}
  \city{Singapore}
  \country{Singapore}
}

\author{Dan Lin}
\affiliation{%
  \institution{Sun Yat-Sen University}
  \city{Guangzhou}
  \country{China}
}

\author{Xingtong Yu}
\affiliation{%
  \institution{The Chinese University of Hong
Kong}
  \city{Hong Kong}
  \country{China}
}

\author{Tat-Seng Chua}
\affiliation{%
  \institution{National University of Singapore}
  \city{Singapore}
  \country{Singapore}
}

\author{Zhiming Zheng}
\affiliation{%
  \institution{Beihang University}
  \city{Beijing}
  \country{China}
}








\renewcommand{\shortauthors}{Miao et al.}

\begin{abstract}
As cross-chain interoperability advances, decentralized finance (DeFi) protocols enable illicit funds to be reorganized into uniform liquid assets that flow throughout the cryptocurrency market. Such operations can bypass monitoring targeted at individual blockchains and thereby weaken current regulatory frameworks.
%
Motivated by these, we introduce UniDetect, a multi-chain cryptocurrency fraud account detection method based on large language models (LLMs).
Specifically, we use domain knowledge to guide the LLM to generate general transaction summary texts applicable to heterogeneous blockchain accounts, which serve as evidence for fraud account detection. Furthermore, we introduce a two-stage alternating training strategy to continuously and dynamically enhance the multimodal joint reasoning for detecting fraudulent accounts based on both the textual evidence and the transaction graph patterns.
Experiments on multiple blockchains show that UniDetect outperforms existing methods 5.57$\sim$7.58\% in Kolmogorov–Smirnov (KS). For cross-chain zero-shot detection, UniDetect identifies over 94.58\% of fraudulent accounts. It also generalizes well to non-blockchain data, delivering a 6.06\% improvement in $F_1$ over existing methods. 
The dataset and source code are available in \url{https://github.com/msy0513/UniDetect}.
\end{abstract}




\maketitle

\section{Introduction}
Web3 envisions a “distributed value internet" that operates without trusted third parties, leveraging blockchain technology for value transfer and programmable transaction logic. Propelled by this vision of decentralization and user empowerment, users actively initiate on-chain transactions, fueling the rapid expansion of the cryptocurrency digital economy \cite{web3}. 
However, the pseudonymous nature of the underlying blockchain technology enables users to bypass “Know Your Customer” (KYC) processes, creating fertile ground for malicious actors and leading to frequent security incidents. According to CertiK\footnote{https://www.certik.com}, total fraud losses in the first half of 2025 exceeded USD 2.47 billion, surpassing the 2024 total.
More recently, as cross-asset and cross-chain transactions have become routine for users, fraudsters have begun exploiting regulatory gaps between blockchains to launder illicit funds.
This trend is further evidenced by industry reports \cite{elliptic2025chainhopping}, which reveal that nearly 20\% of complex illicit cases now span more than ten blockchains, with the total volume of cross-chain criminal activity exceeding USD 21.8 billion in 2025.

\begin{figure}[!t]
    \centering
    \includegraphics[width=0.9\linewidth]{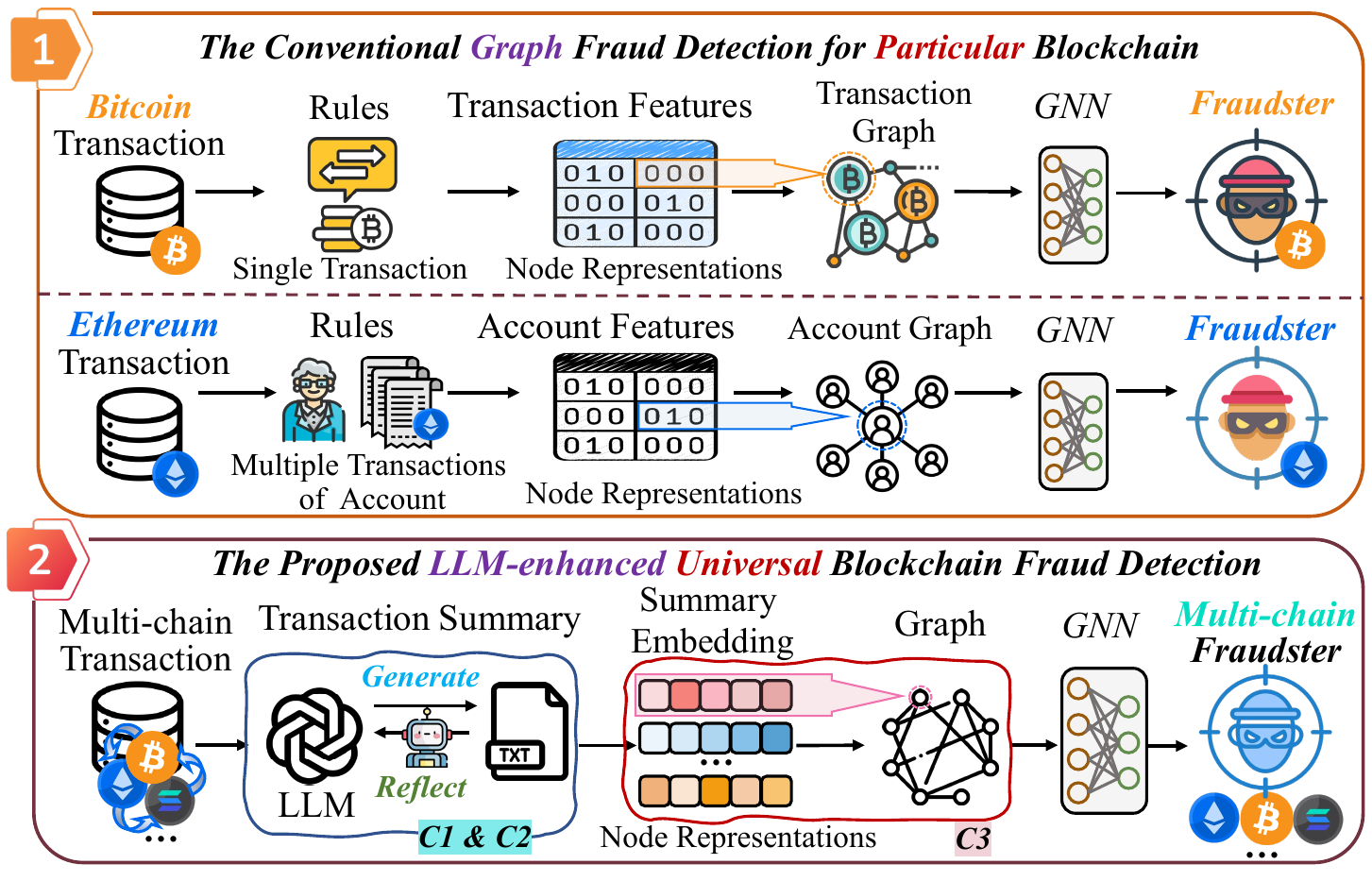}
    \caption{A workflow comparison between chain-specific blockchain fraud account detection methods and our multi-chain universal fraud detection method. Conventional methods rely on chain-specific feature engineering, while our method leverages an LLM-based agent to enable consistent fraud detection across diverse blockchains.
    }
    \label{fig1}
\end{figure}

Existing cryptocurrency fraud detection studies~\cite{I2BGNN,Ethident,Tegdetector,msy,msy2} are primarily designed for a single blockchain, which limits their semantic modelling capabilities and makes it difficult to scale to multi-chain scenarios. Specifically, as illustrated in Figure~\ref{fig1} (1), these methods rely heavily on expert-defined numerical statistical features tailored to specific blockchains and exhibit limited semantic expressiveness. In fact, blockchain transaction data is inherently multimodal: it not only contains structured information such as transfer amounts and frequencies, but also rich unstructured semantic content, such as the natural-language information embedded in the input data message fields of Ethereum~\cite{IDM}. However, existing methods are constrained to single-modality analysis, and the form of data modalities varies considerably across different blockchains. Furthermore, due to substantial differences in data structures, transaction mechanisms, and event semantics across blockchains, current methods struggle to address the multimodal heterogeneous fraud account analysis required in complex cross-chain transaction scenarios.

Therefore, to improve the effectiveness and generalizability of fraud account detection, new paradigms are needed to address the limitations. Large language models (LLMs), with their rich knowledge and strong semantic reasoning capabilities, offer promising potential for extracting deeper insights from complex on-chain transactions~\cite{risktagger}. 
Nevertheless, adapting LLMs for blockchain fraud account detection poses several fundamental \textbf{challenges (\textbf{\textit{C}})} that need to be addressed:
\textbf{\textit{(C1)} Understanding heterogeneous block-chains.} To overcome the shallow feature design and single-chain constraints of existing approaches, LLMs must automatically process and align these disparate data to enable unified multi-chain understanding.
\textbf{\textit{(C2)} Hallucination and over-inference risks.} 
When processing large volumes of transactions, LLMs may generate inaccurate or fabricated information, known as hallucination. In fraud detection scenarios, preventing over-inference and false positives is critical to ensuring reliable, explainable detection results.
%
%
%
\textbf{\textit{(C3)} Effectively integrating multimodal data.}
Numerical features and textual information in transactions provide complementary evidence for understanding transaction semantics. Meanwhile, transaction graphs capture the behavioural patterns of accounts, which are crucial for account identification. 
Although numerical features and textual information in transactions are relatively easy to handle, directly serializing large-scale transaction graphs into text for large language models incurs substantial computational overhead. 
Therefore, it is imperative to develop an efficient graph--language integration framework that supports effective fusion and collaborative modeling of multiple transaction data modalities under explicit computational constraints.

%

To address the above challenges, we propose \textbf{UniDetect}, an LLM-driven cryptocurrency fraud account \textbf{\underline{detect}}ion method designed to improve \textbf{\underline{uni}}versality and effectiveness across heterogeneous blockchains, as illustrated in Figure~\ref{fig1} (2). 
Specifically, (i) for \textbf{\textit{C1}}, we develop a forensic analysis agent, which can utilize LLM’s inherent knowledge and relevant prompts to handle transactions on different blockchains adaptively, and save historical transaction summaries of accounts as evidence for reasoning.
(ii) For \textbf{\textit{C2}}, we elaborately design two summary analysis agents and fine-tune their underlying LLMs via reinforcement learning (RL) to guide the transaction summaries in eliminating irrelevant hallucination and over-inference content unrelated to the task.
(iii) For \textbf{\textit{C3}}, during transaction graph construction, we adopt a semantic-guided sampling and structural compression to reduce computational complexity. In addition, we treat the high-quality transaction summaries, after removing irrelevant information, as accounts’ semantic features and use them to further train the graph model. Together with the preceding fine-tuning stage, this process forms a two-stage alternating training that enables achieving efficient integration of multimodal data.

Our main \textbf{contributions} are summarised as follows:
\begin{itemize}[leftmargin=*, topsep=4pt, partopsep=0pt]
\item We make the first attempt to integrate LLMs with customized fine-tuning strategies to develop a universal detection method that identifies fraudulent accounts across heterogeneous blockchains.

\item Our UniDetect leverages LLM-generated transaction summaries as node features to enhance transaction-graph representation learning, and further employs a two-stage alternating training strategy to jointly refine summary quality while enabling tightly coupled and reliable integration with downstream graph-reasoning tasks.


\item Extensive evaluations on three real cryptocurrency datasets show that UniDetect improves fraud detection performance by up to 8.56\% in $F_1$ and at least 5.57\% in Kolmogorov–Smirnov (KS). Moreover, our method successfully identifies over 94.58\% of real fraudulent accounts in zero-shot cross-chain detection and achieves 4.47\% AUC improvement on cross-domain datasets. 

\end{itemize}

\section{Related work}
We review related work on blockchain-based cryptocurrency fraud detection from three key perspectives: 

\par\noindent
\textbf{General Graph-based Learning Methods.}
These methods typically model transaction behaviours between accounts as graphs and perform account representation learning using classical graph learning models, such as DeepWalk \cite{1deepwalk}, Node2Vec \cite{2node2vec}, GCN \cite{4gcn}, and GAT \cite{4gcn}, or general graph anomaly detection approaches, including CARE-GNN \cite{care-gnn}, BWGNN \cite{BWGNN}, and GRIT \cite{15grit}. However, these methods lack integration of blockchain-specific background knowledge, making them not directly applicable.

\par\noindent
\textbf{Blockchain-specific Detection Methods.}
This line of research incorporates blockchain-specific fraud characteristics into graph learning frameworks to improve detection performance.
For example, Trans2Vec~\cite{Trans2vec} integrates transaction amounts and timestamps into graph-based random walks to enhance traditional random walk models.
Ethident~\cite{Ethident} designs 49 graph augmentation strategies and 16 manually crafted features tailored for Ethereum accounts to identify optimal augmentation and initialization settings.
TEGDetector~\cite{Tegdetector} converts transaction sequences into multiple temporal slices and jointly learns account representations by combining time-series modelling with graph learning techniques.
TSGN~\cite{tsgn} reformulates blockchain fraud detection as a transaction-centric graph classification problem, enabling the capture of significant transaction patterns. However, these methods are applicable only to specific blockchains and cannot generalize to solve complex multi-chain or even cross-chain scenarios.

\par\noindent
\textbf{Multi-chain Detection Methods.}
This line of work aims to reduce reliance on blockchain-specific designs by adopting unified graph reconstruction or sequence-based modelling of transaction behaviours.
SIGTRAN \cite{sigtran} models transactions from different block-chains using a generic graph representation and performs node classification by combining multiple types of features.
BERT4ETH \cite{bert4eth} employs a Transformer \cite{transformer} architecture to encode historical transaction sequences of accounts, enabling chain-independent representation learning.
DIAM \cite{DIAM} constructs directed multi-graphs and applies a gated recurrent unit (GRU) \cite{gru} temporal encoder to initialize account representations without requiring chain-specific features. However, these methods generally separate domain knowledge from the reasoning learning process. This decoupled learning model makes it difficult to effectively understand the semantics of fraudulent accounts.

\begin{figure*}[t]
    \centering
    \includegraphics[width=\linewidth]{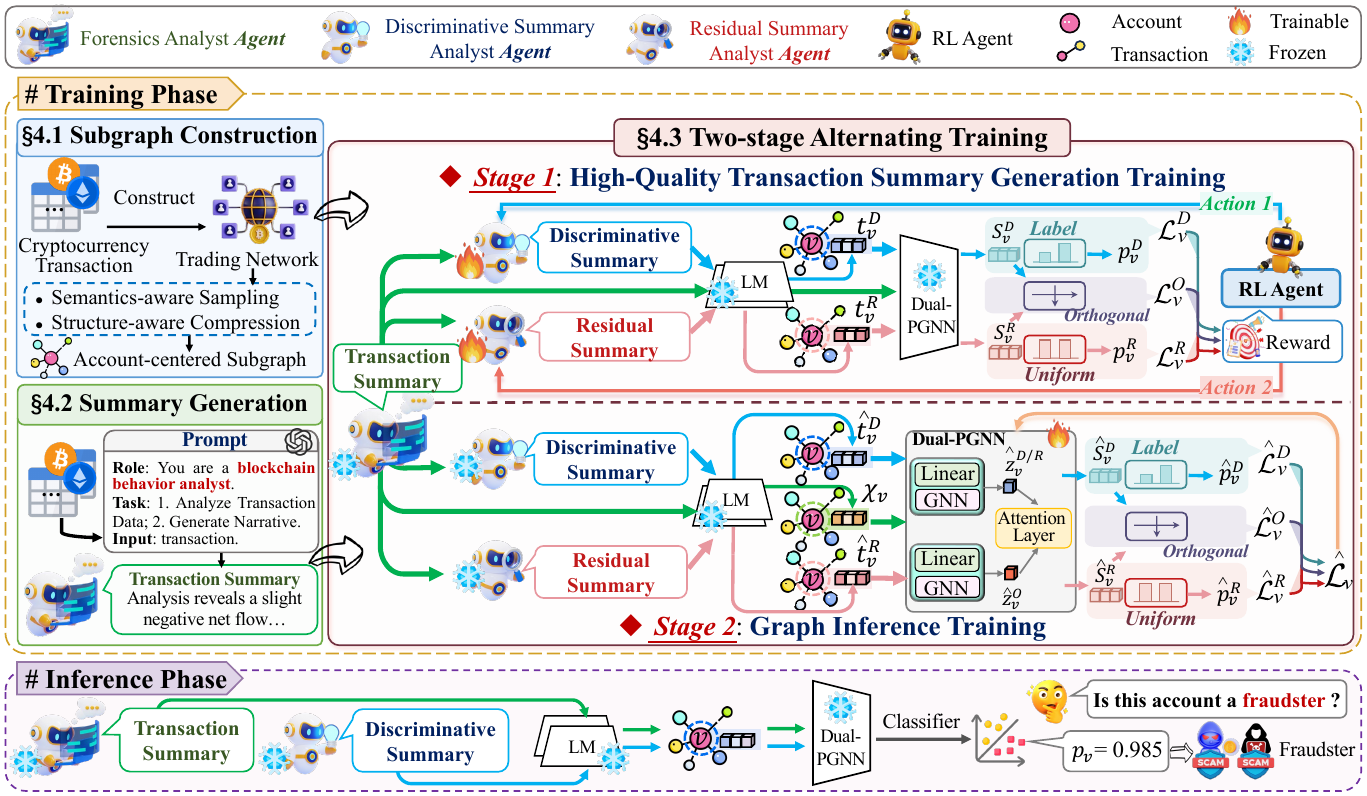}
    \caption{
\textbf{The overall framework of UniDetect}.
During \textbf{training}, we construct account-centered subgraphs, generate transaction summaries via a forensic analysis agent, and apply a two-stage alternating training strategy that jointly fine-tunes LLM-based summary analyst agents and the graph encoder. 
During \textbf{inference}, the forensic and discriminative summary agents collaboratively enrich account representations for fraud prediction.
    }
    \label{fig2}
\end{figure*}

\section{Preliminaries}
\textbf{Data Model}. We represent the transaction graph as \(G = (V, E)\), which consists of the following components:
(i) node set \(V = \{u, v, \ldots\}\) containing \(n\) nodes;
(ii) directed edge set \(E\) containing \(m\) transaction edges; and
(iii) edge attribute matrix \(\mathbf{X}_E \in \mathbb{R}^{m \times d}\), where each row is a \(d\)-dimensional vector encoding the corresponding transaction details.

\par\noindent
\textbf{Problem Definition}. We formulate the fraud account detection task as the node classification. Let \( Y_L \) denote the set of node labels, and
the objective is to learn a binary classifier
$f : G = (V, E, \mathbf{X}_E, Y_L) \longmapsto  Y_U,$
which predicts the labels \(Y_U\) of the unlabeled nodes.

\section{Proposed Approach}
Figure~\ref{fig2} presents the overall framework of UniDetect, which mainly consists of a training phase and an inference phase.
%

\subsection{Subgraph Construction}
Due to the large volume of account transactions, processing the entire transaction graph directly would incur significant time and computational costs. 
Therefore, we first optimize the graph size by sampling the account-centered subgraph and structure-aware graph compression.
%

\par\noindent
\textbf{Semantics-aware Sampling}.
We select a labelled account node $v$ as the centre node and perform $h$-hop neighbour sampling to construct an account-centred transaction subgraph, prioritizing neighbours that exhibit high-value behaviours with the central account. 
Specifically, we sample the top-$K$ most important neighbours at each hop based on their average transaction value. The sampling process can be described as follows:
\begin{equation}
    C_h \;=\; \bigcup_{v \in C_{h-1}} \operatorname{topK}\!\big(\mathcal{N}_v,\; K,\; \mathbf{R}[v,\mathcal{N}_v]\big),
\end{equation}
where $C_h$ is the set of nodes obtained through $h$-th hop sampling ($h = 0$ indicates node $v$), $\mathcal{N}_v$ is the set of 1-hop neighbour nodes centered at node $v$, $K$ is the number of neighbours selected for each hop, $\mathbf{R}[v,\mathcal{N}_v]$ represents the average transaction value of candidate interactions, serving as guidance for selecting adjacent nodes. ${\mathcal{C}_v=\cup_{k=0}^hC_k}$ represents the set of nodes constructed by sampling ${h}$-hop neighbors from the target account $v$. 
If the sample contains duplicate average transaction amounts, we will select the transaction information based on the total transaction value.

\par\noindent
\textbf{Structure-aware Compression}.
Although the subgraph is preliminarily reduced via sampling, its node count may still be of order $K^h$. Therefore, we propose a \underline{s}tructural \underline{i}mportance \underline{g}raph \underline{c}ompression (SIGC) algorithm that removes non-critical nodes while preserving the core structure and connectivity of the subgraph, effectively reducing its size without losing essential information. 
%
\circled{1} \textit{\textbf{Computing the Structural Importance of Neighbour Nodes.}}
We define $S_u$  as the structural importance of neighbour node $u$ in the subgraph:
\begin{align}
    S_u = \frac{\log(a_u^{\text{in}} + a_u^{\text{out}} + 1) + \beta \cdot \log(d_u^{\text{in}} + d_u^{\text{out}} + 1)}{L_u + 1},
\end{align}
where $a_u^{in}$/$a_u^{out}$ denote the inflow/outflow transaction amounts of the neighbour node $u$. $d_u^{in}$/$d_u^{out}$ represents its in/out-degree within the subgraph. $L_u$ is the shortest path length from the central node $v$ to the the neighbour node $u$, computed via Breadth-First Search (BFS) \cite{BFS}. And $\beta$ controls the trade-off between behavioural information and structural signals when computing the node importance score.
\circled{2} \textit{\textbf{Structural Importance Graph Compression.}}
If the $h$-hop transaction subgraph centred on the node $v$ contains more than $N_c$ nodes, we sort the neighbour nodes by their structural importance score $S_u$ and retain the top $N_c$. To preserve connectivity, we trace the shortest paths from the node $v$ to the retained nodes and include all nodes and edges on these paths in the final compressed subgraph.

\subsection{Transaction Summary Generation}
\label{section3.3}
%
After graph construction, we employ an LLM-based forensic analysis agent to automatically process the transactions associated with each account node $v$ and generate an account-level summary $r_v$. By leveraging LLMs’ embedded knowledge and a carefully designed chain-of-thought (CoT) prompting strategy, the agent can consistently handle blockchain accounts with heterogeneous data structures. 
Figure \ref{fig6} illustrates the prompt template, with more details provided in \textbf{Appendix 1}.
%
\begin{figure}[t]
    \centering
    \includegraphics[width=0.98\linewidth]{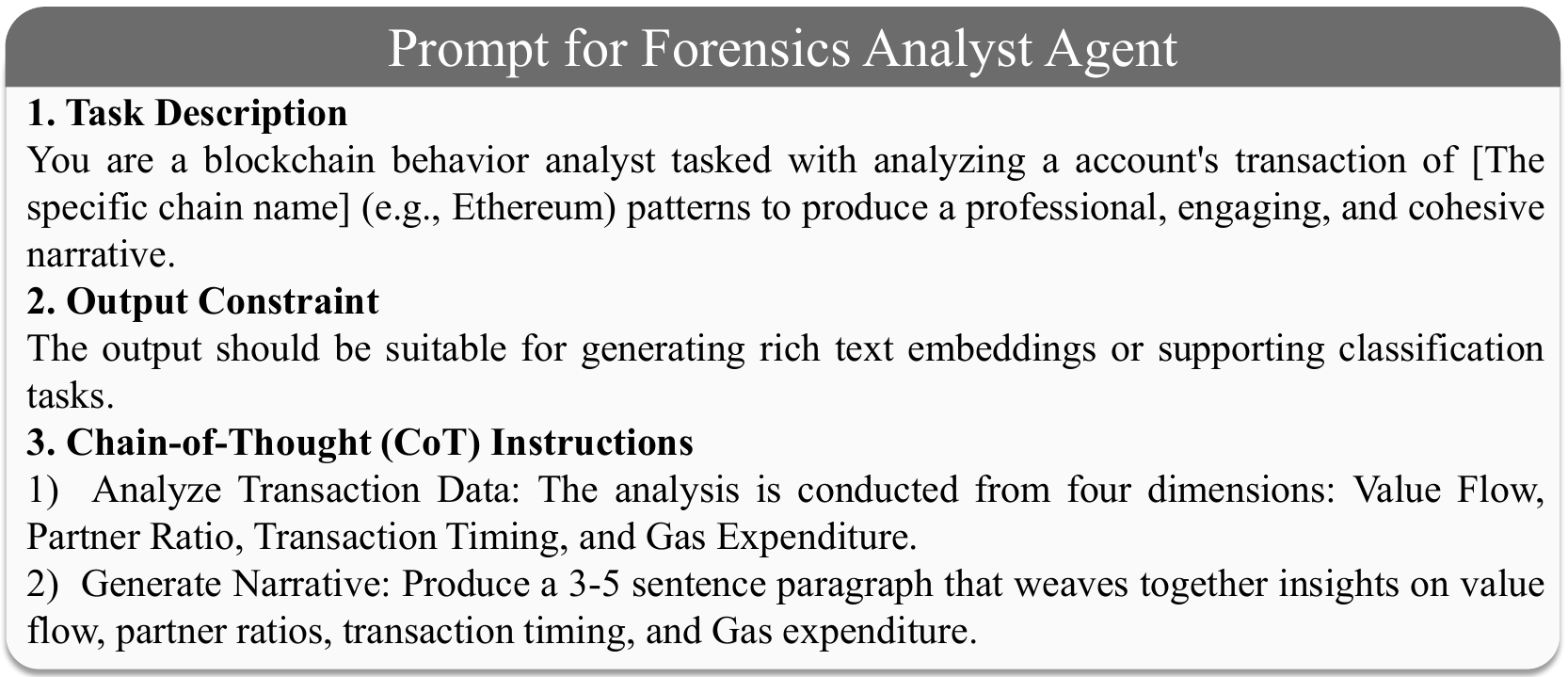}
    \caption{Prompt template of forensics analyst agent.}
    \label{fig6}
\end{figure}

\subsection{Two-stage Alternating Training}


To improve the quality of transaction summaries, we develop two summary analysis agents that share the same LLM backbone. The \textbf{discriminative summary analyst agent} focuses on classification-relevant cues (e.g., value movements, transaction patterns, and log semantics), while the \textbf{residual summary analyst agent} preserves descriptive content that is weakly correlated with detection. 
Then, we adopt a two-stage training strategy to improve training stability and convergence efficiency.

\par\noindent
\noindent$\bullet$ 
\textbf{\textit{Stage {1}: High-Quality Transaction Summary Generation Training:}}

In \textit{Stage 1} (\textit{outer loop}), the parameters of the LLM-based summary analyst agent are fine-tuned while the graph encoder remains frozen. To mitigate the high computational cost of full-parameter LLM fine-tuning, we use Low-Rank Adaptation (LoRA) \cite {LoRA} to improve efficiency. The detailed process is as follows:

\par
\begin{figure}[t]
    \centering
    \includegraphics[width=0.98\linewidth]{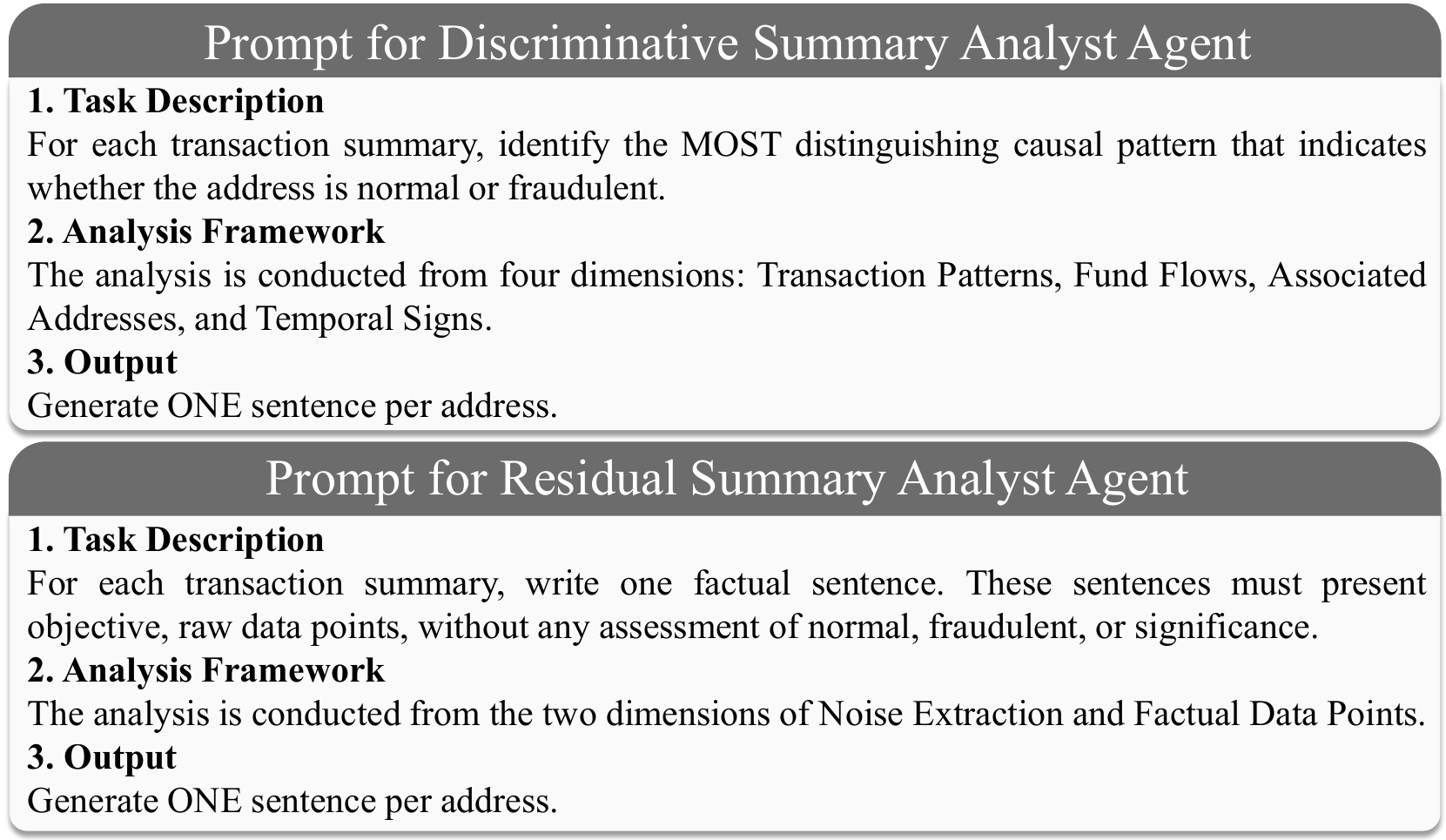}
    \caption{Prompt template of discriminative and residual summary analyst agents.}
    \label{fig7}
\end{figure}

\noindent
\textbf{Discriminative And Residual Summary Generation}.
We leverage a discriminative summary analyst agent to extract discriminative information $r_v^D$ from account transaction summaries $r_v$. 
%
%
In addition, to improve the quality of the discriminative summary, we introduce an auxiliary task that extracts a complementary residual summary $r_v^R$ from the transaction summary $r_v$. 
Figure \ref{fig7} shows the prompts for the discriminative summary and residual summary analyst agents, with further details provided in \textbf{Appendix 1}. The \ul{\textbf{discriminative summary analyst agent}} extracts task-relevant, discriminative information from the initial transaction summary to generate a discriminative transaction summary. Its analysis focuses on the following four aspects:
\textit{\textbf{(1) Transaction patterns:}} extracting transaction behaviour descriptions related to fraudulent activities, such as high-frequency or dense transfers within short time intervals;
\textit{\textbf{(2) Fund flows:}} identifying aggregation–dispersion patterns, for example, funds collected from multiple sources and subsequently transferred through multi-hop paths to multiple destinations;
\textit{\textbf{(3) Associated addresses:}} extracting interactions with known high-risk entities, such as mixers, darknet services, or other risky addresses;
\textit{\textbf{(4) Temporal signs:}} identifying time-related indicators of potential risk, such as transactions concentrated in specific high-risk periods.

In addition, the \underline{\textbf{residual summary analyst agent}} aims to separate statement-like content that is not directly relevant to the fraud identification task from the initial transaction summary. Such information is inherently non-discriminatory and therefore cannot be used to effectively support fraud classification. Its analysis mainly includes two aspects:
\textit{\textbf{(1) Noise extraction:}} identifying and extracting task-irrelevant noise, including expressions arising from model hallucinations or over-reasoning, such as prompt-like language, redundant descriptions, or subjective risk judgments lacking explicit transactional evidence;
\textit{\textbf{(2) Factual statements:}} extracting purely factual descriptions that directly state transaction data but do not convey risk-related characteristics, such as an account’s transaction activity lasting for $X$ months or transaction volumes remaining stable over time.

Subsequently, a fixed-parameter language model (LM) is employed to encode both independently, yielding vector representations $t_v^D$ and $t_v^R$ that serve as initial node features for the graph encoder. Additionally, we feed the original transaction summary $r_v$ from Subsection \ref{section3.3} into the same LM to obtain its representation $\chi_v$.

\par\noindent
\textbf{Dual-path graph neural network (Dual-PGNN) Encoder}. 
The discriminative/residual summary representations $t_v^D$/$t_v^R$ and the original summary representation $\chi_v$ are respectively input into the Dual-PGNN. After encoding, we obtain node representations $S_v^{D}$ and $S_v^{R}$. A detailed introduction to Dual-PGNN is provided in \textit{Stage 2}.

\par\noindent
\textbf{Tri-View Loss.} 
We define the model loss from three perspectives:

\par\noindent
 \underline{i) Discriminative summary loss}: For each central node $v$ of the subgraph, we compute binary cross-entropy (BCE) loss $L_v^D$ between the node’s prediction $p_v^D$ obtained by encoding the discriminative summary as the initial node feature and its ground-truth label $y_v$.
For node $v$, the discriminative summary loss is defined as:
\begin{equation}
\mathcal{L}_v^D = -\sum_{v \in V} [y_v \log(p_v^D) + (1 - y_v) \log(1 - p_v^D)].
\end{equation}
\underline{(ii) Residual summary loss}:
We minimize a Kullback–Leibler (KL) divergence, where the residual summary loss for node $v$ is defined as:
\begin{equation}
    \resizebox{.89\linewidth}{!}{$
            \displaystyle
\mathcal{L}_v^R = -\sum_{v \in V} D_{KL}(p_v^R \| U_v) 
= \sum_{v \in V} \sum_{k \in \{0,1\}} p_v^R(k) \log \frac{p_v^R(k)}{U_v(k)}
        $},
\end{equation}%
where $p_v^R$ denotes the predicted probability obtained by encoding the residual summary as the initial node feature, while $U_v$ represents a uniformly distributed prediction ($U_v = \frac{1}{K}$, where $K$ is the number of label classes).
\underline{(iii) Orthogonal Loss}:
To clearly distinguish node representations from the discriminative and residual summaries $S_v^D$ and $S_v^R$, we impose an orthogonality regularizer between them:
\begin{equation}
\mathcal{L}_v^O = \left\| S_v^D \cdot S_v^R \right\|_2^2,
\end{equation}
where $\cdot$ denotes the dot product used to enforce orthogonality between the two representations, preventing mutual interference in account node embeddings.
Finally, the overall loss of account $v$ consists of the above three terms and can be expressed as:
\begin{equation}
\mathcal{L}_v = \mathcal{L}_v^D + \lambda_1 \cdot \mathcal{L}_v^R + \lambda_2 \cdot \mathcal{L}_v^O,
\end{equation}
where $\lambda_1$ and $\lambda_2$ are hyperparameters.

\par\noindent
\textbf{RL-based Fine-tuning Strategy.}
We fine-tune the LLM using sequence-level REINFORCE~\cite{REINFORCE} with an exponential moving average baseline to improve the discriminative transaction summary.
Specifically, the tri-view loss $\mathcal{L}_v$ of the account node $v$ is converted into a reward:
\begin{equation}
R_v = \exp(-\mathcal{L}_v).
\end{equation}
%
Then, we further compute an advantage by subtracting a moving-average baseline:
\begin{equation}
\hat{A}_v = R_v - b_t,
\end{equation}
where $b_t$ is an exponential moving-average baseline at training step $t$.
%
Finally, the RL objective of node $v$ is defined as:
\begin{equation}
\mathcal{L}^{\mathrm{RL}}_v
=
-\, \hat{A}_v \cdot
\Big(
\log p_v^{\mathrm{Disc}}
+
\log p_v^{\mathrm{Resi}}
\Big),
\end{equation}
where $p_v^{\mathrm{Disc}}$ and $p_v^{\mathrm{Resi}}$ denote the sequence-level generation probabilities of the discriminative and residual summary analyst agent, respectively. 

\par\noindent
\noindent$\bullet$ 
\textbf{\textit{Stage {2}: Graph Inference Training:}}
\label{graph_encoder}

In \textit{Stage 2} (\textit{inner loop}), the LLM is frozen, and the graph encoder begins to train. 
%
Similar to \textit{Stage 1}, the only difference is that we use the two fine-tuned summary analyst agents to generate the discrimination and residual transaction summaries, which we encode as $\hat{t}_v^D$ and $\hat{t}_v^R$. Then, we use the \textbf{Dual-PGNN} to encode the account representation. Specifically, the original transaction summaries $\chi_v$ and discriminative/residual summaries $\hat{t}_v^D$/$\hat{t}_v^R$ are fed into the two branches of the Dual-PGNN. Each branch comprises a backbone GNN and a linear projection that aggregates neighbourhood information while preserving the original summary. The node representation of account $v$ obtained by encoding the discriminative/residual summary as node features is:
\begin{equation}
\hat{Z}_v^{D/R} = \mathrm{GNN}(\hat{r}_v^{D/R}, \mathcal{A}_v) + \theta(\hat{r}_v^{D/R}),
\end{equation}
and the node representation obtained by the original summary is:
\begin{equation}
\hat{Z}_v^{O} = \mathrm{GNN}(\chi_v, \mathcal{A}_v) + \theta(\chi_v),
\end{equation}
where $\theta(\cdot)$ is a linear layer and $ \mathcal{A}_v$ denotes the adjacency matrix of the subgraph centered at node $v$. Then, an attention layer performs weighted fusion of the two branch representations to obtain the final node embedding:
\begin{equation}
\hat{S}_v^{D/R} = W_1 \cdot \hat{Z}_v^{D/R} + W_2 \cdot \hat{Z}_v^{O},
\end{equation}
where $W_1$ and $W_2$ are learnable parameters derived from attention weights, which adjust the relative importance of the discriminative/residual summary and the original transaction summary during fusion.
Finally, we employ the tri-view loss to optimize the Dual-PGNN parameters.

\subsection{Validation and Test}
During verification and testing, we feed the original transaction summaries generated by the forensic analysis agent and the discriminative summaries generated by the discriminative summary analysis agent into the two branches of the dual-PGNN, respectively. The encoded and aggregated account node representations are then input into a fully connected (FC) layer to predict the probability of fraud.

\section{Experiments}
This section aims to answer the following research questions: \textbf{RQ1}: How does UniDetect perform in heterogeneous blockchain transactions?
\textbf{RQ2}: What is the contribution of each component of UniDetect?
\textbf{RQ3}: How well does UniDetect generalize?
\textbf{RQ4}: How sensitive is UniDetect to its hyperparameters?
\textbf{RQ5}: What is the computational overhead of UniDetect in practice? Finally, we analyzed the rationality and effectiveness of our method by examining a real fraud account detection case.

\subsection{Experimental Settings}

\textbf{Datasets:}
We evaluate UniDetect on Ethereum and Bitcoin, two representative public blockchains with fundamentally different data structures, as shown in Table \ref{tab1}.
(1) \textit{\textbf{Ethereum:}} 
We obtain the fraud labels from the Etherscan label cloud\footnote{\url{https://etherscan.io/}}, XLabelCloud\footnote{\url{https://xblock.pro/\#/labelcloud}} and CryptoScamDB\footnote{\url{https://cryptoscamdb.org/scams}}. 
These labels cover major illegal account types such as phishing, Upbit vulnerability attacks, gambling, and honeypots. Additionally, we consider special identity accounts, such as miners and financial service providers, as benign because they are known to be non-fraudulent. 
Then, we use the Etherscan API\footnote{\url{https://etherscan.io/}} to collect historical transaction data of both fraudulent and non-fraudulent accounts from “2015-08-07" to “2025-04-10”. 
During data preprocessing, we remove transactions with zero amounts and failed transactions, and then construct the Ethereum transaction graph. 
(2) \textit{\textbf{Bitcoin:}} We utilise two publicly available datasets: Bitcoin-M and Bitcoin-L \cite{DIAM}, with varying node and edge scales to evaluate model performance comprehensively. 
In addition, to evaluate our method’s generalization to other fields, we include the \textit{\textbf{Instagram}} \cite{Instagram} social network dataset. All datasets are split into training, validation, and test sets following an 8:1:1 ratio.
Further details of the datasets and graph construction are provided in \textbf{Appendix 2}.

\begin{table}[htbp]
    \centering
    \setlength{\tabcolsep}{10pt} 
    \caption{Statistics of the blockchain datasets.}
    \resizebox{\linewidth}{!}{
    \renewcommand{\arraystretch}{1} 
    \begin{tabular}{l|r|r|r}
        \toprule
        Dataset & Ethereum & Bitcoin-M & Bitcoin-L  \\
        \midrule
        \#Nodes   & 187,960 & 2,505,841 & 20,085,231 \\
        \#Edges   & 397,038 & 14,181,316 & 203,419,765 \\
        \#Fraud   & 1,991 & 46,930 & 362,391  \\
        \#Normal  & 1,439 & 213,026 & 1,271,556 \\
        \#Fraud (\%) & 1.06 & 1.87 & 1.80  \\
        \bottomrule
    \end{tabular}
    }
    \label{tab1}
\end{table}


\noindent\textbf{Evaluation metrics:}
We use four common metrics: Precision, Recall, $F_1$, and AUC. In addition, we use the KS, a widely used metric in credit scoring, to measure the separability between benign and fraudulent accounts, defined as $\mathrm{KS} = \max|F_1(x) - F_2(x)|$.
For all metrics, higher values indicate better performance.


\noindent \textbf{Baselines:}
We compare UniDetect against 20 baseline methods, which can be grouped into four major categories: \textit{\textbf{(1) Graph representation learning methods}}: DeepWalk \cite{1deepwalk}, Node2Vec \cite{2node2vec}, GCN \cite{4gcn}, GAT \cite{7GAT}, GIN \cite{5gin}, GraphSAGE \cite{6graphsage}, and APPNP \cite{APPNP};
\textit{\textbf{(2) Graph anomaly detection method}}: CARE-GNN \cite{care-gnn}, BWGNN \cite{BWGNN}, GRIT \cite{15grit}, DGA-GNN \cite{Dga-gnn}, PMP \cite{PMP}, FLAG \cite{FLAG};
\textit{\textbf{(3) Single-chain fraud detection methods}}: Trans2Vec \cite{Trans2vec}, TSGN \cite{tsgn}, Ethident \cite{Ethident}, and TEGDetector \cite{Tegdetector};
\textit{\textbf{(4) Multi-chain fraud detection methods}}: SIGTRAN \cite{sigtran}, BERT4ETH \cite{bert4eth} and DIMA \cite{DIAM}. 
Details are provided in \textbf{Appendix 3}.

\noindent\textbf{Implementation Details:}
We initiate a 2-hop neighbour sampling, retaining up to $K=10$ neighbours per hop. In the compression process, the hyperparameter $\beta = 2$ is used to prioritize aggregating highly correlated nodes, and the final subgraph size $N_c$ is set to 10. In the two-stage alternating training, we use 
gemma-3-4b-it \cite{gemma} as the base LLM for all agents. Sentence-BERT \cite{Sentence-BERT} serves as the LM for generating embeddings. The Dual-PGNN is configured with two layers, a hidden-layer size of 64, and a GNN backbone of CARE-GNN. 
In the tri-view loss function, we set $\lambda_1$ to 0.05 and $\lambda_2$ to 0.3, respectively.
Furthermore, we utilize the AdamW optimizer with a fixed learning rate of 5e-6 and adopt early stopping to prevent overfitting. The training process involves 2 outer epochs and 10 inner epochs. 
All experiments are conducted on a machine with an Intel(R) Xeon(R) Gold 6248R CPU (3.00 GHz), 128 GB of RAM, and two NVIDIA A6000 GPUs. 
Detailed baseline implementation settings are provided in \textbf{Appendix 4}. In addition, to prevent the LLM from exploiting pre-trained knowledge of known labels, we strictly exclude account identifiers and label information from both the input data and prompt instructions. A detailed discussion is provided in \textbf{Appendix 5 (Q1)}.

\subsection{Overall Effectiveness (RQ1)}

Tables \ref{t2} and \ref{t3} present results from two representative blockchain platforms, including the average metrics and their standard deviations across multiple runs.
The \textbf{\best{best}} and the \underline{\second{second}} are marked. UniDetect$^\dagger$ uses the LLM in a zero-shot setting without fine-tuning. A detailed analysis of the RL training convergence is provided in \textbf{Appendix 5 (Q2)}, confirming the stability of the proposed fine-tuning process.
Several key conclusions can be drawn:

\begin{itemize}[leftmargin=*]
    \item UniDetect consistently outperforms all competing methods across all evaluation metrics and datasets.
    \item Among graph representation learning and graph anomaly detection methods are constrained by challenges in real-world cryptocurrency fraud detection, including severe label imbalance and limited transaction information.
    \item Single-chain blockchain methods rely on manual feature engineering and fail to capture key transaction patterns, leading to limited performance.
    \item Multi-chain methods typically treat transactions as direction-aware numeric sequences for encoding learning. However, it often struggles to fully capture the structural semantics of the transaction graph; moreover, separating the sequence encoder and the graph reasoning also makes it difficult to consistently capture the semantics of fraudulent accounts.
\end{itemize}

%

\begin{table}[!t]
    \centering
    \setlength{\tabcolsep}{5.5pt}
    \caption{Evaluation on Ethereum dataset (in percentage \%). The \textbf{\best{best}} and the \underline{\second{second}} are marked. UniDetect$^\dagger$ uses the LLM in a zero-shot setting without fine-tuning.}
        \resizebox{\linewidth}{!}{
        \begin{tabular}{l|c|c|c|c|c}
        \toprule
        \multicolumn{6}{c}{\textbf{Ethereum}} \\
        \midrule
        Model  & Precision & Recall & $F_1$ & {AUC} & KS \\
        \midrule
        DeepWalk     & 71.91$_{0.56}$ & 52.89$_{1.14}$ & 60.95$_{0.31}$ & 64.82$_{0.89}$ & 42.31$_{0.37}$\\
        Node2Vec     & \underline{\second{82.68}}$_{0.23}$ & 41.77$_{0.01}$ & 55.60$_{0.49}$ & 60.15$_{0.26}$ & 36.75$_{1.12}$ \\
        GCN          & 40.97$_{1.17}$ & 50.00$_{0.49}$ & 45.04$_{0.73}$ & 42.37$_{1.67}$ & 26.39$_{0.92}$\\
        GAT          & 40.97$_{1.19}$ & 50.00$_{1.33}$ & 45.04$_{0.86}$ & 42.34$_{0.32}$ & 26.33$_{0.14}$\\
        GIN          & 50.00$_{0.05}$ & 45.03$_{0.51}$ & 47.39$_{0.16}$ & 64.91$_{0.03}$ & 29.12$_{0.67}$\\
        GraphSAGE    &70.12$_{0.63}$  & 79.07$_{1.73}$ & 74.27$_{1.34}$ & 76.28$_{1.42}$ & 44.23$_{0.83}$\\
        APPNP        & 59.88$_{0.93}$ & 51.08$_{1.47}$ & 55.04$_{0.59}$ & 52.74$_{0.16}$ & 31.41$_{1.87}$\\
        \midrule
        CARE-GNN     & 69.58$_{0.12}$ & 81.56$_{0.41}$ & 75.05$_{1.15}$ & 56.71$_{1.35}$ & 38.46$_{0.03}$\\ 
        BWGNN        & 80.33$_{0.02}$& 82.21$_{1.43}$ & 81.26$_{0.74}$ & 81.14$_{0.72}$ & 50.73$_{1.18}$\\ 
        GRIT         & 55.89$_{1.35}$ & 56.41$_{1.03}$ & 56.15$_{1.36}$ & 61.63$_{1.18}$ & 42.03$_{1.28}$\\
        DGA-GNN      & 82.43$_{1.12}$ & 82.91$_{0.71}$ & 82.67$_{0.91}$ & 81.39$_{0.94}$ & \underline{\second{61.09}}$_{0.48}$\\ 
        PMP          & 67.00$_{0.51}$ & 81.86$_{0.54}$ & 73.49$_{0.48}$ & 71.79$_{0.73}$ & 37.60$_{0.64}$\\  
        \midrule
        Trans2Vec    & 79.47$_{0.61}$ & 48.44$_{0.66}$ & 60.19$_{0.65}$ & 73.95$_{0.28}$ & 42.51$_{1.19}$\\
        TSGN        & 73.75$_{1.28}$ & 71.30$_{1.12}$ & 72.50$_{1.83}$ & 75.44$_{0.03}$ & 43.68$_{0.68}$\\
        Ethident     & 80.28$_{0.55}$ & 87.80$_{0.52}$ & 83.86$_{1.57}$ & 82.27$_{0.55}$ & 60.17$_{1.66}$\\
        TEGDetector  & 81.33$_{0.49}$ & 80.42$_{0.75}$ & 80.86$_{1.30}$ & 79.33$_{0.74}$ & 59.42$_{0.38}$\\
        \midrule
        SIGTRAN      & 79.75$_{1.18}$ & \underline{\second{88.55}}$_{0.62}$ & 83.89$_{0.31}$ & 79.71$_{0.28}$ & 50.43$_{0.65}$\\
        BERT4ETH     & 81.65$_{0.78}$ & 85.62$_{0.44}$ & 83.59$_{1.76}$ & 81.82$_{0.25}$ & 58.77$_{1.03}$\\
        DIAM         &81.96$_{0.13}$ &86.38$_{0.63}$	&\underline{\second{84.09}}$_{0.42}$	&\underline{\second{82.67}}$_{0.11}$ & 60.23$_{0.17}$\\
        \midrule
        UniDetect$^\dagger$    & 85.14$_{0.43}$ & 85.63$_{0.36}$ & 85.38$_{0.45}$ & 81.82$_{1.23}$ & 63.32$_{0.39}$ \\
        UniDetect    &\makecell{\textbf{\best{91.39}} }$_{0.60}$	&\makecell{\textbf{\best{93.95}} }$_{0.56}$	&\makecell{\textbf{\best{92.65}} }$_{0.34}$	&\makecell{\textbf{\best{94.82}}}$_{0.24}$ & \makecell{\textbf{\best{68.94}}}$_{0.12}$\\ 
        \midrule
        \rowcolor{blue!12}
        Improv.\textcolor{red}{\small$\triangle$}     &+8.71  &+5.40 &+8.56 &+12.15  &+7.58\\
        \bottomrule
        \end{tabular}
    }
\label{t2}
\end{table}

\begin{table*}[t]
    \centering
    \setlength{\tabcolsep}{8pt}
    \caption{Evaluation on Bitcoin-M and Bitcoin-L datasets (in percentage \%). The notations are the same as those in Table \ref{t2}.}
    \resizebox{\linewidth}{!}{%
    \begin{tabular}{l!{\vrule width 0.4pt}ccccc!{\vrule width 0.4pt}ccccc}
    \toprule
    \multicolumn{1}{l}{\textbf{Dataset}}& \multicolumn{5}{c}{\textbf{Bitcoin-M}}
    & \multicolumn{5}{c}{\textbf{Bitcoin-L}} \\
    \midrule
    \textbf{Model}  & Precision & Recall & $F_1$ & {AUC} & {KS} & Precision & Recall & $F_1$ & {AUC} & {KS}\\
    \midrule
    DeepWalk     & 63.42$_{0.65}$  & 57.68$_{0.95}$ & 60.42$_{0.74}$ & 69.51$_{0.31}$ & 35.65$_{0.94}$ &62.31$_{1.59}$ &60.63$_{0.31}$ &61.45$_{1.48}$  &67.00$_{2.11}$  &32.62$_{0.16}$ \\
    Node2Vec     & 76.81$_{0.59}$ & 72.61$_{2.71}$ & 74.63$_{2.34}$ & 68.72$_{1.59}$ & 45.63$_{0.36}$  &68.49$_{0.59}$ &60.75$_{1.94}$ &64.39$_{0.68}$ &64.46$_{1.35}$  &36.02$_{0.64}$\\
    GCN          & 79.90$_{0.01}$ & 81.21$_{0.26}$ & 80.49$_{0.96}$ & 78.33$_{1.80}$ & 46.71$_{0.36}$ & 70.11$_{1.94}$ &73.35$_{0.01}$ &71.69$_{0.73}$ &73.72$_{1.02}$ &43.88$_{0.79}$\\
    GAT          & 86.16$_{0.46}$ & 81.45$_{1.62}$ & 83.71$_{1.47}$ & 81.27$_{0.87}$ & 53.42$_{0.70}$ & 79.45$_{0.54}$ &65.73$_{0.61}$ &71.80$_{1.33}$ &80.44$_{1.25}$ &44.79$_{0.43}$ \\
    GIN          & 64.68$_{0.73}$ & 61.88$_{0.04}$ & 63.14$_{0.01}$ & 77.17$_{0.42}$ & 38.67$_{1.68}$  & 70.06$_{0.87}$ &55.45$_{1.38}$ &61.70$_{1.75}$ &74.27$_{0.66}$ &43.68$_{0.74}$\\
    GraphSAGE    &87.17$_{1.12}$  & 83.27$_{1.38}$ & 85.17$_{2.25}$ & 83.28$_{0.96}$  & 58.26$_{1.72}$ & 73.16$_{1.28}$ &74.79$_{2.06}$ &74.00$_{0.76}$ &89.93$_{1.35}$ &45.43$_{1.73}$\\
    APPNP        & 83.27$_{1.03}$ & 79.58$_{2.33}$ & 81.36$_{0.75}$ & 80.23$_{1.23}$ & 51.33$_{1.20}$ & 71.69$_{1.76}$ &70.34$_{0.76}$ &71.00$_{0.98}$ &83.79$_{0.29}$ & 44.17$_{2.23}$\\
    \midrule
    CARE-GNN  & 86.69$_{2.31}$ & 86.58$_{0.01}$ & 86.67$_{1.85}$ & 85.04$_{1.89}$ & 60.22$_{3.16}$ & 64.61$_{0.05}$ &80.38$_{1.48}$ &71.62$_{0.01}$ &70.33$_{1.67}$ &37.23$_{1.61}$  \\  
    BWGNN & 86.09$_{1.62}$ & 87.12$_{0.01}$ & 86.66$_{2.10}$ & 81.55$_{1.25}$ & 60.60$_{1.68}$ & 71.15$_{2.97}$ &80.38$_{0.85}$ &75.46$_{1.51}$ &71.57$_{0.55}$& 47.57$_{2.24}$\\ 
    GRIT         & 68.63$_{2.16}$ & 69.41$_{2.42}$ & 69.02$_{2.36}$ & 67.64$_{1.75}$ & 38.31$_{2.08}$  & 65.43$_{1.39}$ &62.78$_{1.46}$ &64.11$_{1.73}$ &64.36$_{2.08}$ & 35.44$_{1.77}$\\
    DGA-GNN      & 87.15$_{0.58}$ & \underline{\second{87.95}}$_{1.74}$ & 87.55$_{0.65}$ & \underline{\second{87.36}}$_{0.95}$ & 64.40$_{0.38}$ & 68.66$_{1.07}$ &80.07$_{1.38}$ &73.97$_{1.59}$ &72.67$_{1.95}$ & 43.19$_{0.02}$\\ 
    PMP     & 85.16$_{1.53}$ & 86.30$_{1.54}$ & 85.73$_{2.43}$ & 85.73$_{0.01}$ & 64.61$_{2.49}$ & 59.31$_{0.01}$ &80.38$_{1.29}$ &71.64$_{0.46}$ &70.42$_{1.40}$ & 44.68$_{2.21}$\\  
    \midrule
    SIGTRAN  & 75.97$_{0.28}$ & 52.24$_{1.45}$ & 61.94$_{1.09}$ & 74.30$_{2.38}$ & 43.82$_{0.72}$ & 80.52$_{0.32}$ & 75.22$_{2.50}$ & 77.80$_{2.16}$ & 80.45$_{0.71}$ &50.32$_{0.72}$ \\
    BERT4ETH    & 80.03$_{2.18}$ & 59.95$_{0.71}$ & 68.55$_{0.42}$ & 78.32$_{0.66}$ &47.11$_{1.07}$  &84.68$_{0.48}$ & 77.31$_{0.01}$ & 80.81$_{2.49}$ & 86.19$_{0.53}$ & 58.11$_{0.01}$\\
    DIAM         &\underline{\second{88.83}}$_{1.04}$ &86.39$_{0.12}$	&\underline{\second{87.57}}$_{1.56}$	&84.10$_{0.52}$ &\underline{\second{64.88}}$_{1.03}$     &\underline{\second{89.58}}$_{0.68}$ &\underline{\second{90.11}}$_{0.64}$	&\underline{\second{89.80}}$_{0.64}$	&\underline{\second{90.05}}$_{0.78}$  & \underline{\second{65.76}}$_{0.83}$   \\
    \midrule
    UniDetect$^\dagger$  & 89.32$_{0.30}$ & 90.16$_{0.66}$ & 89.74$_{0.82}$ & 88.23$_{0.62}$ &67.41$_{0.27}$  &85.23$_{0.01}$ & 77.43$_{0.43}$ & 81.10$_{1.18}$ & 86.39$_{0.52}$ & 60.16$_{0.16}$\\
    UniDetect     &\makecell{\textbf{\best{90.61}}$_{0.48}$ }	&\makecell{\textbf{\best{95.47}$_{0.99}$} }	&\makecell{\textbf{\best{92.98}}$_{0.64}$ }	&\makecell{\textbf{\best{93.83}$_{0.32}$}} &\makecell{\textbf{\best{70.45}$_{1.16}$}} &\makecell{\textbf{\best{90.61}$_{0.50}$} } 	&\makecell{\textbf{\best{96.43}$_{1.16}$} }	&\makecell{\textbf{\best{93.41}$_{0.49}$} }	&\makecell{\textbf{\best{95.22}$_{0.71}$}} &\makecell{\textbf{\best{72.25}$_{0.66}$}}     \\
    \midrule
    \rowcolor{blue!12}
    Improv.\textcolor{red}{\small$\triangle$}    &+1.78  &+7.52 &+5.41 &+6.47 &+5.57  & +1.03 &+6.32 & +3.61 &+5.17 &+6.49\\
    \bottomrule
    \end{tabular}
}
\label{t3}
\end{table*}

\subsection{Ablation Study (RQ2)}
To validate the effectiveness of each key component in UniDetect, we conducted five comprehensive ablation studies on the Ethereum dataset, including the following settings:

\begin{itemize}[leftmargin=*]
    \item \textbf{\textit{w/o} GC} removes the SIGC algorithm.
    \item \textbf{\textit{w/o} TS. train} removes the two-stage alternating training and uses only the origin account summaries as node features.
    \item \textbf{\textit{w/o} Resi. agent} removes the residual summary analyst agent as well as the associated residual summary loss and orthogonal loss.
    \item \textbf{\textit{w/o} RL} removes RL-based fine-tuning and reduces optimization to standard backpropagation.
    \item \textbf{\textit{w/o} Dual-PGNN} removes the Dual-PGNN Encoder and replaces it with a standard GNN, without explicitly preserving self-features as an independent channel during aggregation.
\end{itemize}

The ablation results are shown in Table \ref{t5}, and the main findings are as follows: 
(1) “\textit{w/o} GC” exhibits that compressing the subgraphs appropriately can significantly reduce the time costs with minimal impact on performance.
(2) “\textit{w/o} TS. train” shows degraded performance, demonstrating that improving transaction summary quality is essential to overall detection effectiveness.
(3) “\textit{w/o} Resi. agent” underperforms UniDetect, suggesting that incorporating residual summaries to contrast is important for enhancing summary quality.
(4) “\textit{w/o} RL” leads to noticeable performance degradation, highlighting the importance of RL feedback in optimizing LLM fine-tuning.
(5) “\textit{w/o} Dual-PGNN” results in reduced performance, reflecting that self-feature injection is crucial for mitigating neighbourhood camouflage.
Additionally, experiments on the choice of LLM backbone for the proposed agents are provided in \textbf{Appendix 5 (Q3)}.

\begin{table}[t]
    \centering
    \caption{Ablation studies of UniDetect on the Ethereum dataset. }
    \resizebox{0.8\linewidth}{!}{%
    \begin{tabular}{l|c|c|c}
    \toprule
    Methods   & $F_1$ & {AUC} & Time (s/account) \\
    \midrule
    \textit{w/o} GC       & \textbf{93.62} & 94.14 & 40.37\\
    \textit{w/o} TS. train           & 85.65 & 82.52& / \\
    \textit{w/o} Resi. agent   & 78.37 & 79.81& / \\
    \textit{w/o} RL     & 87.50 & 86.63 & /\\
    \textit{w/o} Dual-PGNN         & 88.50 & 89.63 & /\\
    \midrule
    UniDetect  &92.65	&\textbf{94.82} & \textbf{1.26}\\
    \bottomrule
    \end{tabular}
    }
    \label{t5}
\end{table}

\begin{figure}[t]
\centering
\begin{subfigure}{0.48\linewidth}
    \centering
    \includegraphics[width=\textwidth]{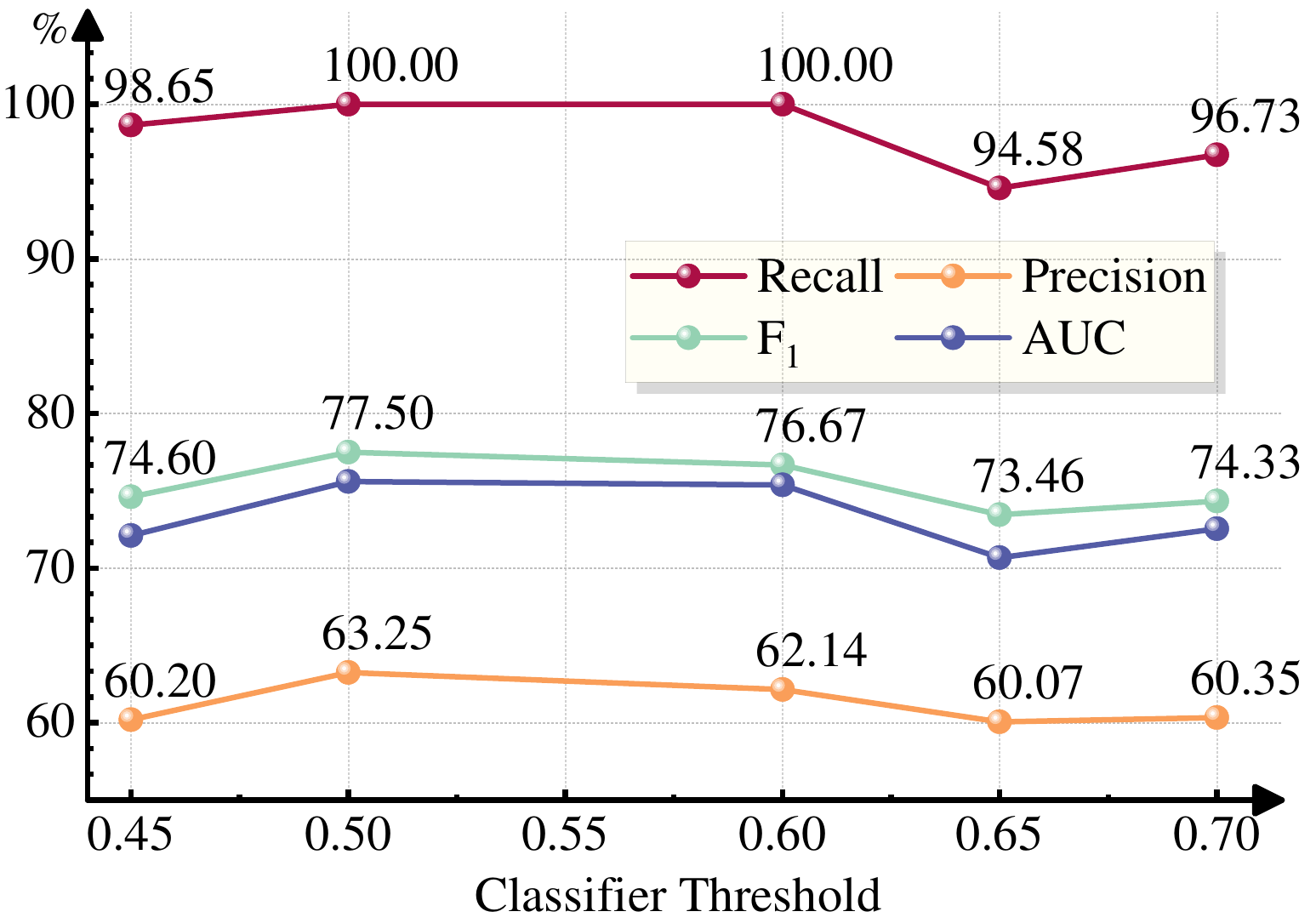}
    \caption{Cross-chain inference.}
    \label{fig:4a}
  \end{subfigure}\hfill
  \begin{subfigure}{0.48\linewidth}
    \centering
    \includegraphics[width=\textwidth]{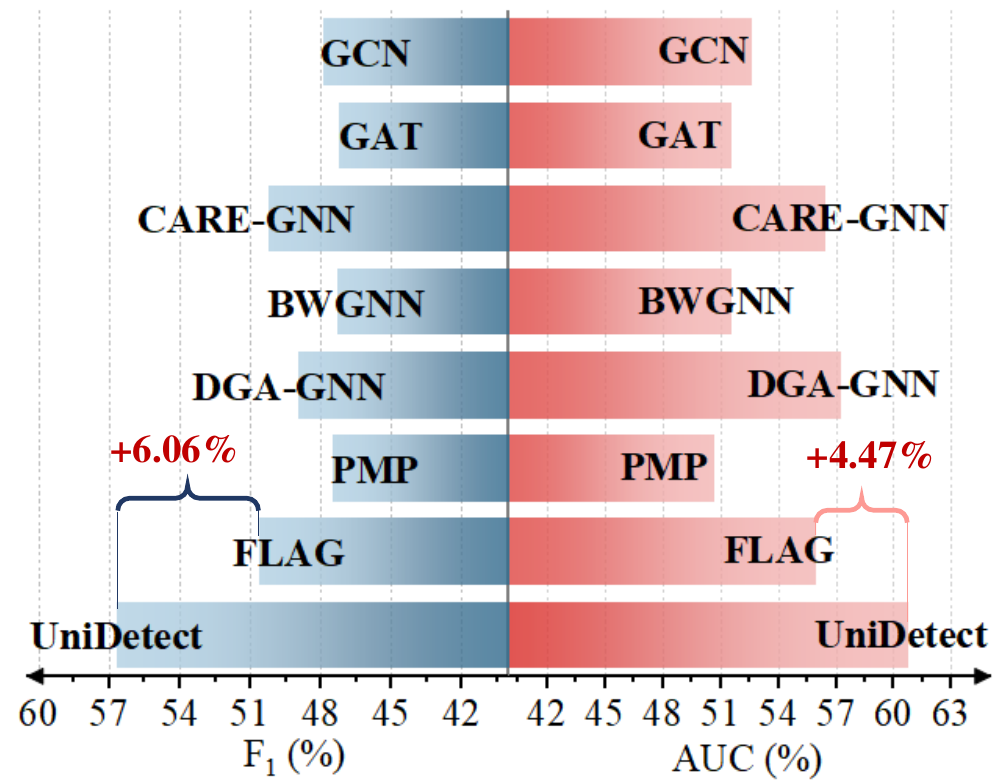}
    \caption{Cross-domain transfer.}
    \label{fig:4b}
  \end{subfigure}
  \caption{Experimental results of UniDetect’s generalization ability. Cross-chain inference (a) shows that fraud detection capability transfers across blockchains with different data structures, while cross-domain transfer (b) demonstrates its effectiveness beyond blockchain scenarios.}
\label{figure4}
\end{figure}
\subsection{Generalization Ability Investigation (RQ3)}
To evaluate the generalization ability of the proposed method, we conducted two experiments, as shown in Figure \ref{figure4}:
(a) \textbf{Cross-chain zero-shot inference}, where UniDetect trained on Ethereum is directly applied to Bitcoin. After extensively varying the classification threshold to mitigate sensitivity, cross-chain zero-shot inference exhibits inherent asymmetry—strong cross-chain generalization for fraudulent accounts but insufficient precision in identifying normal accounts.
%
(b) \textbf{Cross-domain transfer}: we extend UniDetect to the Instagram graph dataset, where the proportion of business users (the minority class) is 36.29\%. The experimental results show that UniDetect remains robust and effective, demonstrating the generality of the proposed approach.
\begin{figure*}[t]
\centering
\begin{subfigure}{0.22\textwidth}
    \centering
    \includegraphics[width=\textwidth]{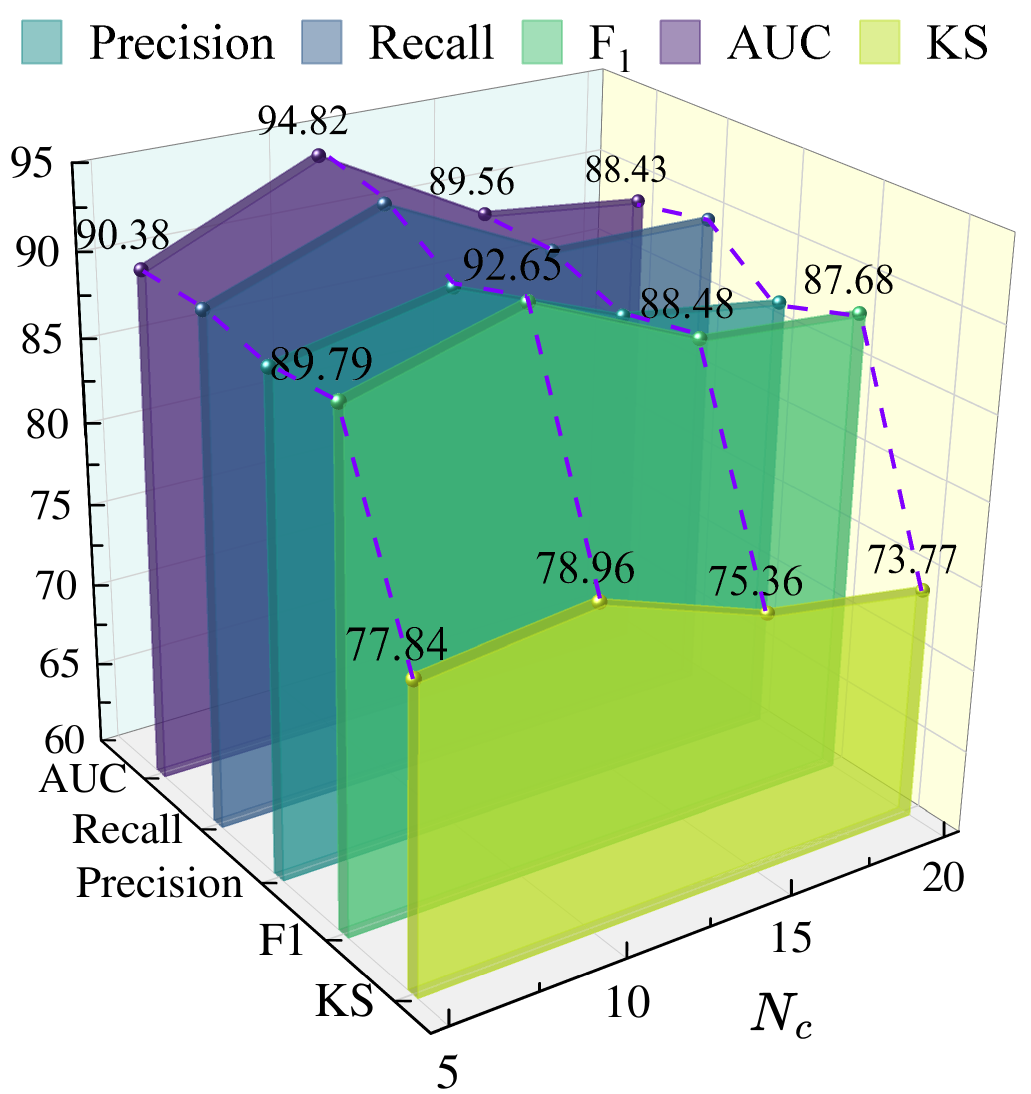}
    \caption{$N_c$.}
    \label{fig:5a}
  \end{subfigure}\hfill
  \begin{subfigure}{0.22\textwidth}
    \centering
    \includegraphics[width=\textwidth]{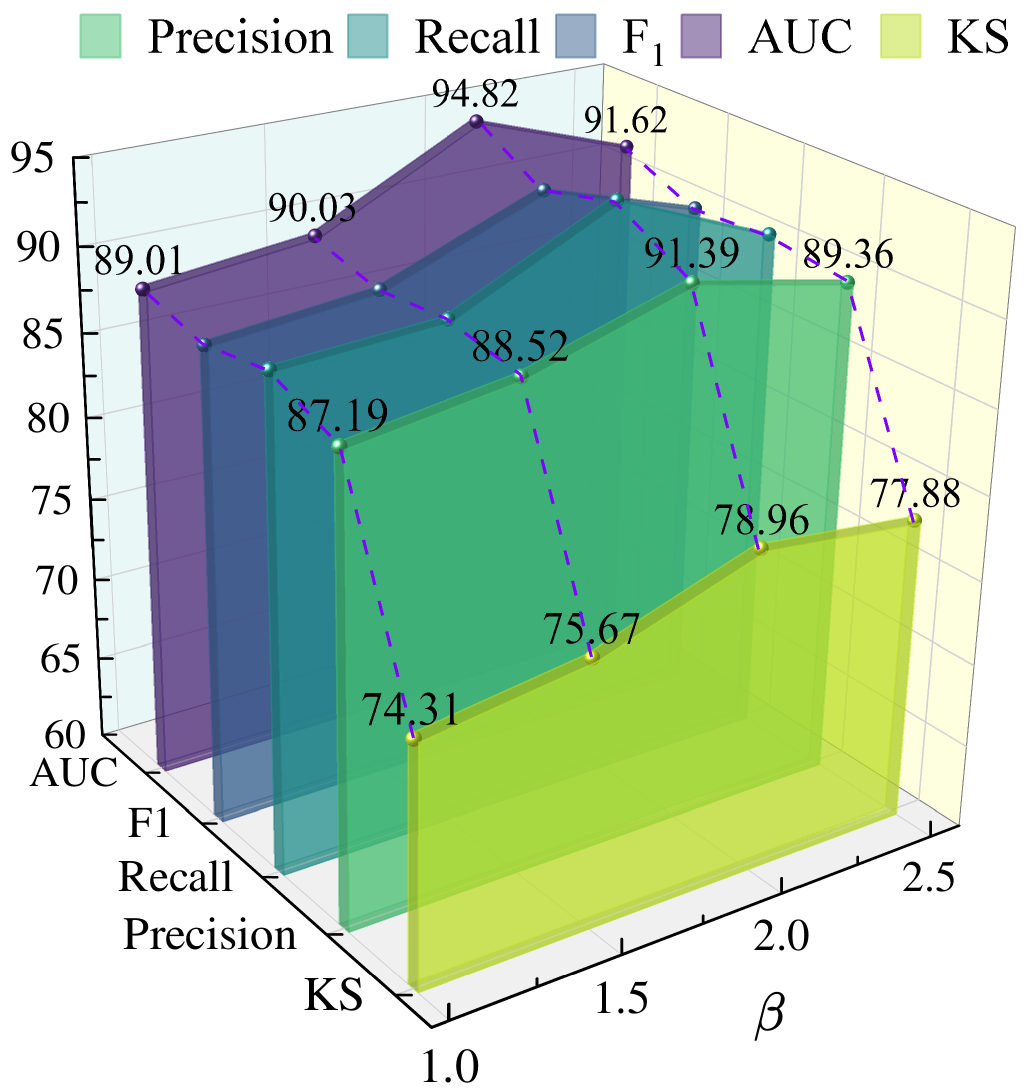}
    \caption{$\beta$.}
    \label{fig:5b}
  \end{subfigure}\hfill
  \begin{subfigure}{0.22\textwidth}
    \centering
    \includegraphics[width=\textwidth]{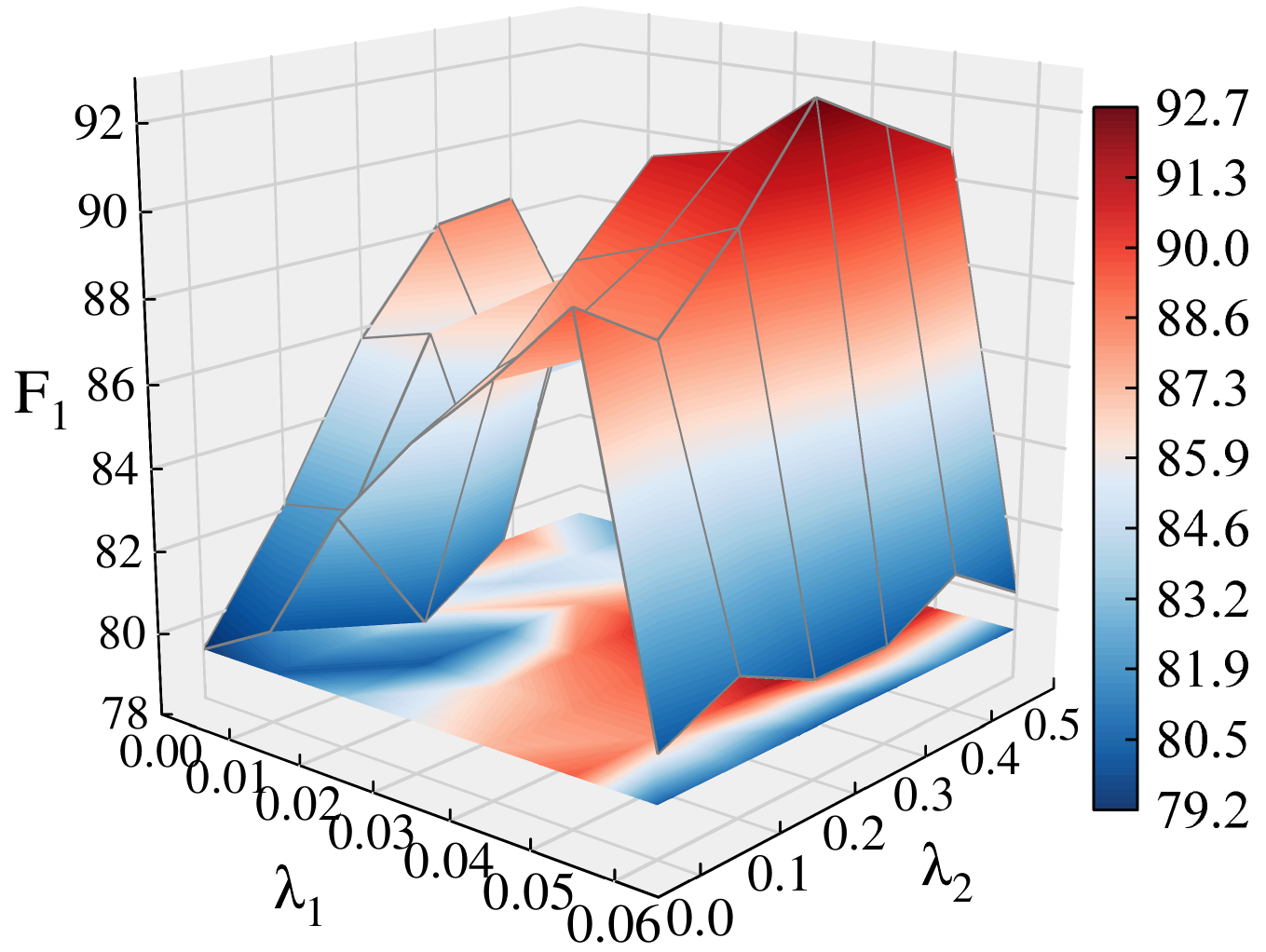}
    \caption{$F_1$ metric under varying $\lambda_1$ and $\lambda_2$.}
    \label{fig:5c}
  \end{subfigure}\hfill
  \begin{subfigure}{0.22\textwidth}
    \centering
    \includegraphics[width=\textwidth]{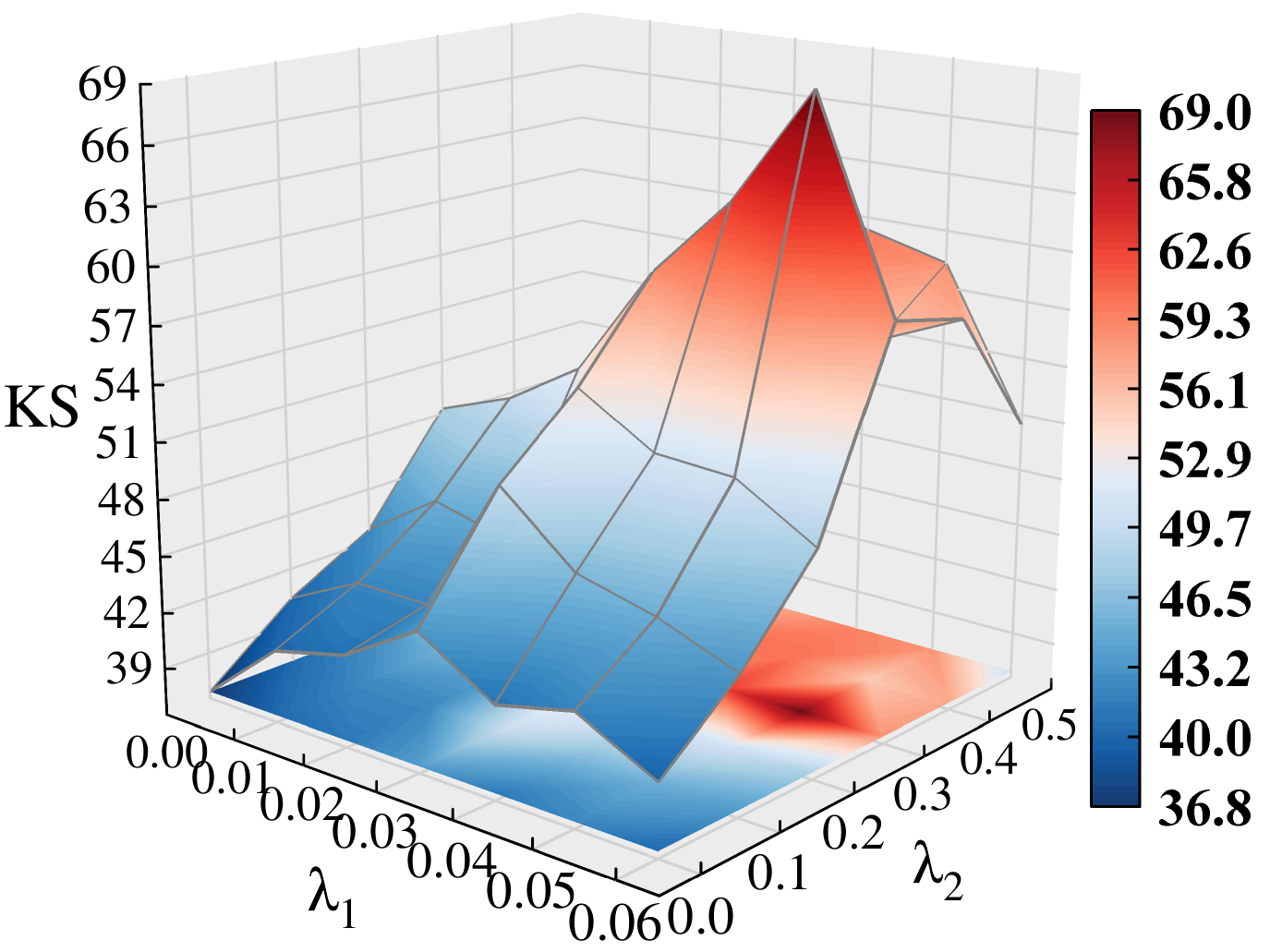}
    \caption{KS metric under varying $\lambda_1$ and $\lambda_2$.}
    \label{fig:5d}
  \end{subfigure}
  \caption{Hyperparameter sensitivity analysis on the Ethereum dataset reveals that UniDetect maintains stable and competitive performance across a broad range of hyperparameter configurations.}
\label{figure5}
\end{figure*}
\subsection{Hyper-parameter Sensitivity Analysis (RQ4)}
Figure \ref{figure5} presents the hyperparameter sensitivity analysis of UniDetect on the Ethereum dataset, focusing on the node threshold $N_c$ for subgraph compression, the signal balancing factor $\beta$ used in neighbourhood structural importance estimation, and the hyperparameter weights used in the tri-view loss function. Three key observations can be drawn:
(1) A larger $N_c$ introduces more benign nodes into the sampled subgraph, weakening the neighbourhood representation of the fraudulent centre node. Reducing $N_c$ alleviates this issue, whereas an overly small $N_c$ leads to insufficient neighbourhood information and degraded performance. 
(2) When $\beta$ is small, the model becomes more sensitive to noise induced by transaction amounts. In contrast, an excessively large $\beta$ overemphasizes structural importance, leading the model to mistakenly treat highly connected nodes (e.g., exchanges or ICO wallets) as important neighbours, thereby distracting the model and reducing its discriminative power.
(3) We analyze the effect of the weighting parameters $\lambda_1$ and $\lambda_2$ in the tri-view loss. Figures \ref{fig:5c} and \ref{fig:5d} show the $F_1$ and KS results under different settings. As $\lambda_1$ and $\lambda_2$ increase, overall performance first improves and then declines, remaining relatively stable across a wide range of values.

\subsection{Computational efficiency (RQ5)}

To evaluate the computational efficiency of UniDetect in practical deployment scenarios, we conduct a comprehensive profiling of graph construction time, model training time, inference time, and model parameters, with results reported in Table~\ref{tab:efficiency}. In the graph construction, the average time for subgraph sampling and compression per account is only 1.26 s. In the training, \textit{Stage 1} and \textit{Stage 2} together take approximately 13 h; although introducing the LLM incurs additional training overhead, this is a necessary cost of leveraging LLM capabilities. In the inference, UniDetect achieves an online inference latency of only 1.47 ms/sample with a throughput of 680 samples/s, comparable to existing lightweight graph-based methods and confirming its practical deployability. Regarding model parameters, thanks to the LoRA fine-tuning strategy, the trainable parameters account for only 0.07\% of the total model parameters. In summary, UniDetect exchanges an acceptable one-time training cost for substantial gains in cross-chain generalisation and detection performance, striking a favourable balance between practicality and effectiveness.

\begin{table}[t]
\centering
\caption{Efficiency analysis of UniDetect on the Ethereum dataset. Despite the additional training overhead introduced by the LLM component, UniDetect remains practical for real-world deployment with acceptable inference latency and throughput.}
\label{tab:efficiency}
\setlength{\tabcolsep}{12pt}
\renewcommand{\arraystretch}{1.15}
\begin{tabular}{lc}
\toprule
\textbf{Component} & \textbf{Cost} \\
\midrule
\rowcolor{gray!15}\multicolumn{2}{l}{\textit{Data processing Time} (s/sample)} \\
\quad Graph construction     & 1.26 \\
\rowcolor{gray!15}\multicolumn{2}{l}{\textit{Training Time} (h)} \\
\quad Stage 1: RL fine-tuning (LLM)   & 12.42 \\
\quad Stage 2: GNN training            & 0.38 \\
\rowcolor{gray!15}\multicolumn{2}{l}{\textit{Inference Time} (ms/sample)} \\
\quad  Inference      & 1.47 \\
\rowcolor{gray!15}\multicolumn{2}{l}{\textit{Throughput} (samples/s)} \\
\quad Inference throughput                 & 680 \\
\rowcolor{gray!15}\multicolumn{2}{l}{\textit{Model parameters (M)}} \\
\quad LLM (Gemma-3-4B-it + LoRA)         & 4303.30 (3.22 trainable) \\
\quad Dual-PGNN                        & 0.04 \\
\bottomrule
\end{tabular}
\end{table}

\subsection{Study Case}

To validate the rationality of our method, we present an analysis from a real-world Ethereum account detection case.

For an account 0xfe68f28599b19c5d8a562e8cc7f7b07c36e0a99d, the forensics analyst agent generates the following \textbf{raw transaction summary}: The target node executed 3,742 transactions involving 81 unique addresses over approximately nine years, receiving inflows totalling 5,454.23 ETH. The collection behaviour exhibits significant net outflow dominance with extensive partner interactions, consistent with large-scale multi-hop transfer activity. Transactions occurred primarily during European business hours. The node has been previously flagged in Europol's 2022 cryptocurrency monitoring report.
Then, the \textbf{discriminative transaction summary retains}:
1. Collection node: significant net outflow dominance with extensive multi-party interactions, suggesting a potential laundering operation. 2. Distribution node: large-scale multi-hop dispersal pattern with sustained high-frequency transaction behaviour.
And \textbf{the residual transaction summary} isolates:
Transactions occurred primarily during European business hours. The node has been previously flagged in Europol's 2022 cryptocurrency monitoring report.


The trained analyst agent is able to distinguish task-relevant content from raw transaction summaries and, more importantly, to identify over-inference and hallucinations introduced by the LLM, such as the timezone inference issue highlighted in the case study. The resulting discriminative summary serves as a transaction history profile for the corresponding account in the transaction graph, which is jointly used with the behavioural topology to determine the account identity.

\section*{CONCLUSION}

In this paper, we make the first attempt to apply LLMs and fine-tuning techniques to multi-chain cryptocurrency fraud account detection and propose UniDetect. 
We construct three LLM-based agents to generate transaction summaries as forensic clues, and then dynamically and closely combine them with the transaction graph learning through a two-stage alternating fine-tuning strategy.
Extensive experiments show that UniDetect consistently outperforms state-of-the-art methods across multiple blockchain datasets and generalizes well to unseen chains and new application domains.

\bibliographystyle{ACM-Reference-Format}
\bibliography{sample-base}










\clearpage
\appendix

\section*{Supplementary Material}
\par\noindent
\textbf{Outline.}
This supplementary material is divided into four sections: Section \ref{section1} introduces the prompt templates for the multiple agents used in this paper. Section \ref{section2} details the dataset construction process and its specific data information. Section \ref{section3} describes the baseline methods. Section \ref{section4} presents the implementation details of the baseline methods in the experiments. Section \ref{section5} addresses common questions regarding training fairness, RL training convergence, and the choice of LLM backbone for the proposed agents.

\section{Templates of agents}
\label{section1}
Figure \ref{fig1} illustrates the prompt templates of the \ul{\textbf{forensics analyst agent}}.
This agent is responsible for analyzing raw transaction records of an account and summarizing its transactional behaviour from four dimensions: value flow, counterparty (partner), transaction timing, and gas expenditure, to generate an initial transaction summary.

Figure \ref{fig2} illustrates the prompt templates for the discriminative summary analyst agent and the residual summary analyst agent, which share a unified role definition.
The \ul{\textbf{discriminative summary analyst agent}} extracts task-relevant, discriminative information from the initial transaction summary to generate a discriminative transaction summary. Its analysis focuses on the following four aspects:
\begin{figure}[!t]
    \centering
    \includegraphics[width=\linewidth]{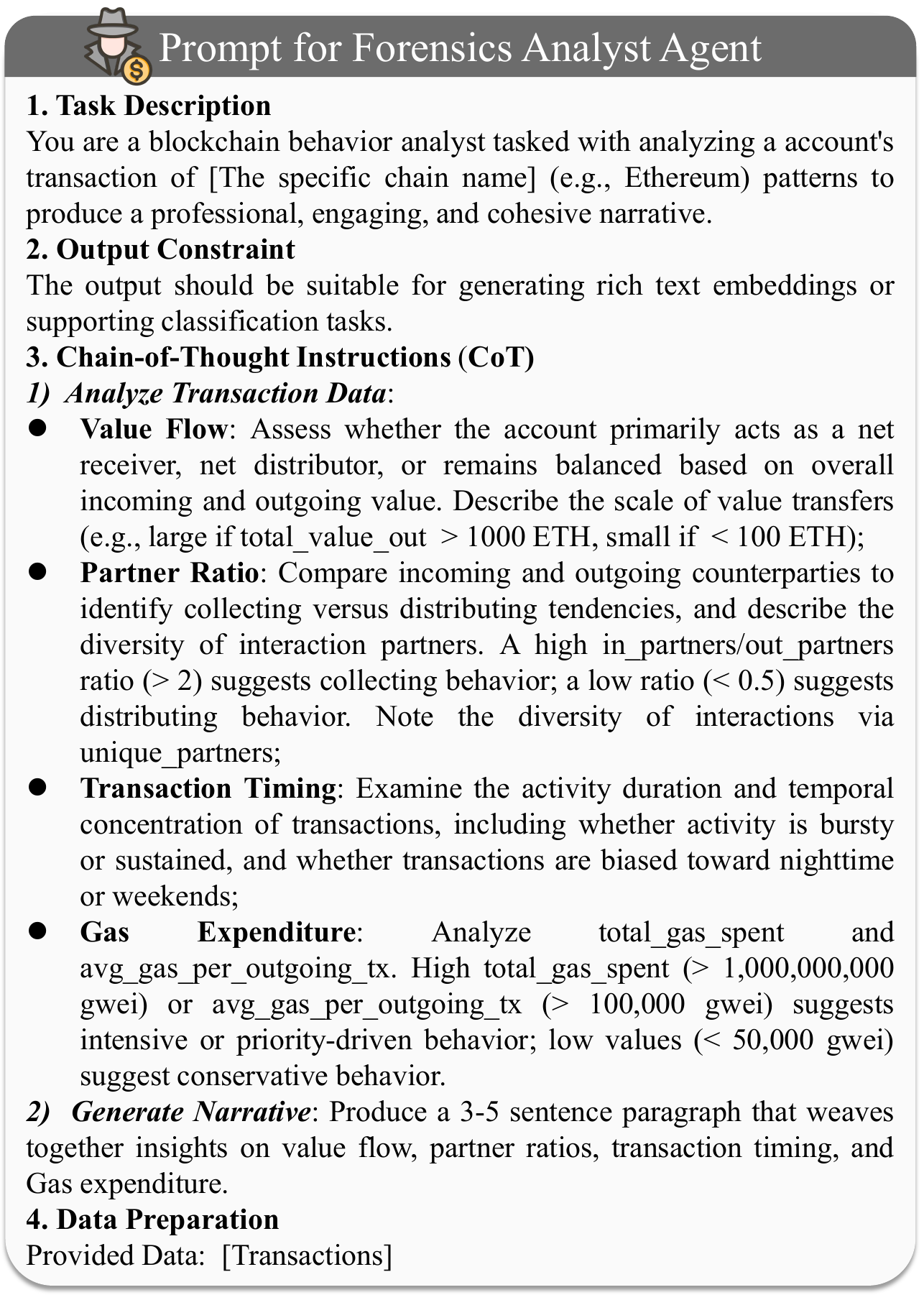}
    \caption{CoT prompt template of the forensics analyst agent.}
    \label{fig1}
\end{figure}
\textit{\textbf{(1) Transaction patterns:}} extracting transaction behaviour descriptions related to fraudulent activities, such as high-frequency or dense transfers within short time intervals;
\textit{\textbf{(2) Fund flows:}} identifying aggregation–dispersion patterns, for example, funds collected from multiple sources and subsequently transferred through multi-hop paths to multiple destinations;
\textit{\textbf{(3) Associated addresses:}} extracting interactions with known high-risk entities, such as mixers, darknet services, or other risky addresses;
\textit{\textbf{(4) Temporal signs:}} identifying time-related indicators of potential risk, such as transactions concentrated in specific high-risk periods.

In addition, the \underline{\textbf{residual summary analyst agent}} aims to separate statement-like content that is not directly relevant to the fraud identification task from the initial transaction summary. Such information is inherently non-discriminatory and therefore cannot be used to effectively support fraud classification. Its analysis mainly includes two aspects:
\textit{\textbf{(1) Noise extraction:}} identifying and extracting task-irrelevant noise, including expressions arising from model hallucinations or over-reasoning, such as prompt-like language, redundant descriptions, or subjective risk judgments lacking explicit transactional evidence;
\textit{\textbf{(2) Factual statements:}} extracting purely factual descriptions that directly state transaction data but do not convey risk-related characteristics, such as an account’s transaction activity lasting for $X$ months or transaction volumes remaining stable over time.

\section{Detailed Descriptions of Dataset.}
\label{section2}
In our experiments, we evaluate UniDetect using Ethereum and Bitcoin, two representative public blockchains with fundamentally different data structures.

\begin{figure}[!t]
    \centering
    \includegraphics[width=\linewidth]{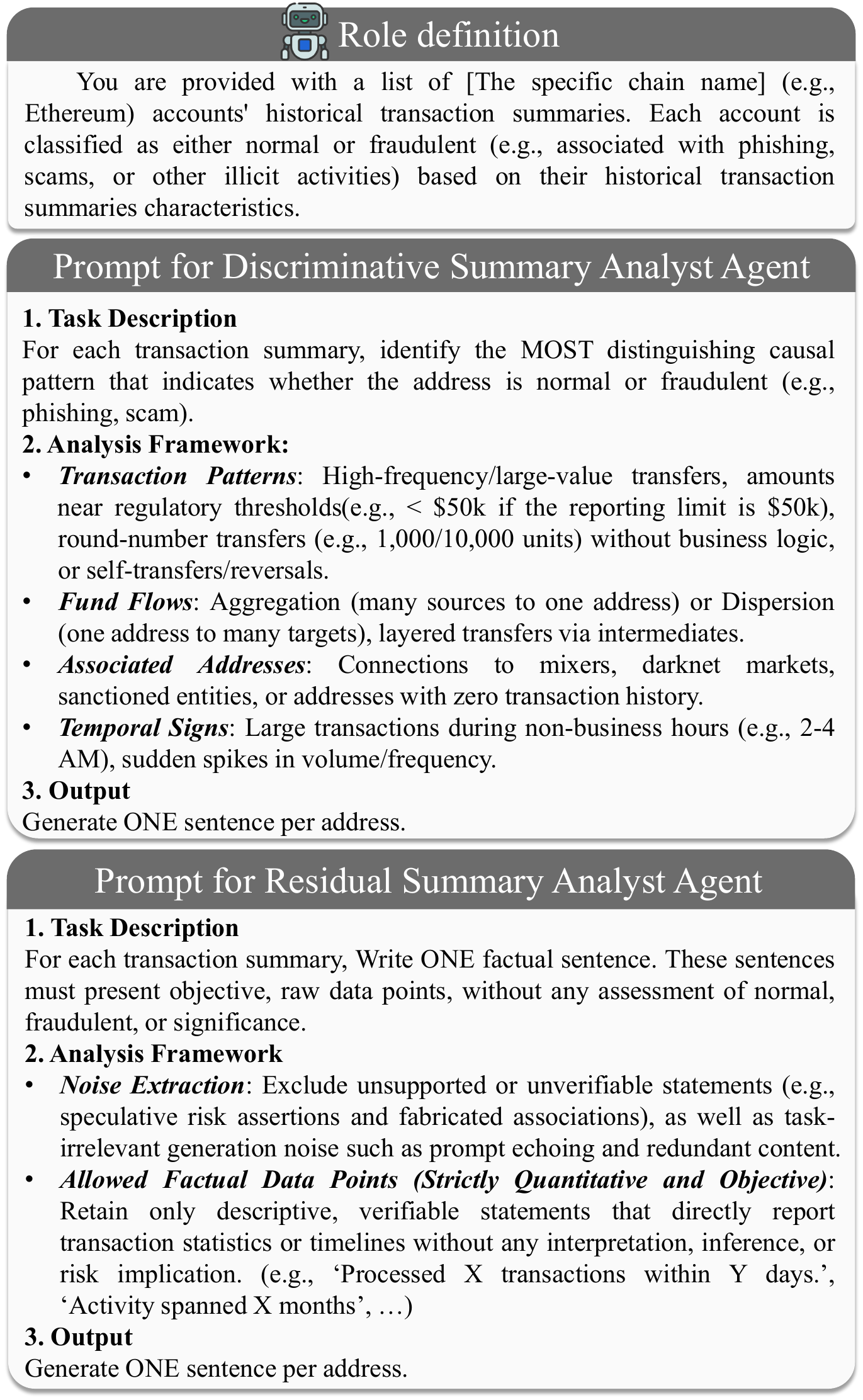}
    \caption{CoT prompt template of discriminative summary analyst agent and residual summary analyst agent.}
    \label{fig2}
\end{figure}

\par\noindent
\textit{\textbf{Ethereum.}}  We obtain the fraud labels from the Etherscan label cloud\footnote{\url{https://etherscan.io/}}, XLabelCloud\footnote{\url{https://xblock.pro/\#/labelcloud}} and CryptoScamDB\footnote{\url{https://cryptoscamdb.org/scams}}. These labels cover major illegal account types such as phishing, Upbit vulnerability attacks, gambling, and honeypots. Additionally, we consider special identity accounts, such as miners and financial service providers, as benign because they are known to be non-fraudulent. Then, we use the Etherscan API\footnote{\url{https://etherscan.io/}} to collect historical transaction data of both fraudulent and non-fraudulent accounts from “2015-08-07" to “2025-04-10”. During data preprocessing, we remove transactions with zero amounts and failed transactions, and then construct the Ethereum transaction graph. 
%
Each Ethereum transaction has a sender and a receiver, and we represent the transfer behavior as directed edges in the graph. The transfer amount and transaction frequency are used as edge attributes. For multiple transfer behaviors between the same pair of accounts, they are uniformly represented as a single edge in the graph, with edge attributes including the cumulative transfer amount and cumulative transaction frequency across all related transactions.

\par\noindent
\textit{\textbf{Bitcoin.}} We leveraged two publicly available datasets: Bitcoin-M and Bitcoin-L \cite{DIAM}, which differ in the numbers of nodes and edges, to enable a more comprehensive evaluation of our model. 
(1) \textbf{Bitcoin-M} contains the first 1.5 million transactions before June 2015.
For each transaction edge in Bitcoin-M, we include four attributes: input value, input amount, output value, and output amount.
(2) \textbf{Bitcoin-L} covers all transactions from June to September 2015. Each transaction edge in Bitcoin-L is associated with five attributes: input value, input amount, output value, output amount, and transaction fee. Following common practice, addresses related to gambling and mixing services—both strongly correlated with money-laundering activities—are labelled as fraudulent, while all remaining accounts are treated as benign. For repeated transfers between the same pair of accounts, we keep only a single edge in the graph and aggregate the corresponding transaction attributes.

In addition, to evaluate our method’s generalization to other fields, we include the \textit{\textbf{Instagram}} \cite{Instagram} social network dataset, and its statistics are provided in Table \ref{tab1}. The Instagram dataset models users as nodes and their follow relationships as edges. Each user's profile description serves as the initial feature representation for the node. Our task is to classify the nodes' account types, distinguishing between business and regular accounts. Similar to the common class imbalance setting in fraud detection tasks, business accounts account for a relatively low proportion in this dataset, so they are treated as the minority class and used as positive examples.

\begin{table}
    \centering
    \setlength{\tabcolsep}{5pt}
    \caption{Statistics of the Instagram datasets.}
    \resizebox{\linewidth}{!}{%
    \begin{tabular}{lrrrrr}
        \toprule
        Dataset  & \#Nodes & \#Edges & \#Pos. & \#Neg. & \#Pos. ratio(\%)\\
        \midrule
        Instagram & 11,339   & 198,448 &  4115 & 7224 & 36.29      \\
        \bottomrule
    \end{tabular}
    } 
    \label{tab1}
\end{table}

\section{Detailed Descriptions of Baselines}
\label{section3}

\begin{itemize}
\item DeepWalk  \cite{1deepwalk} employs random walks to generate node sequences, which are then used to learn node embeddings through a Skip-Gram model.

\item Node2Vec \cite{2node2vec} extends DeepWalk by balancing local and global exploration to learn more nuanced node embeddings.

\item GCN \cite{4gcn} uses graph convolutions to aggregate node features from the neighbourhood for node representation learning.

\item GAT \cite{7GAT} applies attention mechanisms to weigh the importance of neighbouring nodes during feature aggregation.

\item GIN \cite{5gin} leverages the graph isomorphism principle to learn node embeddings that are invariant to graph permutations.

\item GraphSAGE \cite{6graphsage} employs a sampling-and-aggregation framework to generate node embeddings from large graphs efficiently.

\item APPNP \cite{APPNP} approximates personalized PageRank scores to capture node relationships for representation learning.

\item CARE-GNN \cite{care-gnn} mitigates fraudster camouflage with label-aware neighbour sampling and RL-based neighbour selection.

\item BWGNN \cite{BWGNN} introduces beta graph wavelets to construct band-pass spectral filters for detecting anomaly-related high-frequency components.

\item GRIT \cite{15grit} incorporates graph inductive biases into Transformers without message passing via random-walk relative positional encodings and degree information.

\item DGA-GNN \cite{Dga-gnn} boosts fraud detection with decision-tree binning for non-additive features and dynamic neighbour grouping for aggregation.

\item PMP \cite{PMP} distinguishes homophilic and heterophilic neighbours via partitioned message passing for graph fraud detection.

\item FLAG \cite{FLAG} enhances graph-based fraud detection by integrating LLM-extracted textual semantics to mitigate neighbourhood camouflage and improve node discrimination.

\item Trans2Vec \cite{Trans2vec} introduces a transaction-aware network embedding that encodes transfer amounts and temporal information for Ethereum phishing detection.

\item TSGN \cite{tsgn} identifies phishing accounts by modelling transaction subgraphs as attribute-aware transaction subgraph networks that capture temporal and topological transaction patterns.

\item Ethident \cite{Ethident} proposes a hierarchical graph attention framework with contrastive self-supervision to model account interaction graphs for Ethereum account de-anony-mization.

\item TEGDetector \cite{Tegdetector} models phishing detection as a dynamic graph classification task by learning spatial–temporal patterns from transaction evolution graphs with adaptive time-aware attention.

\item SIGTRAN \cite{sigtran} constructs transaction signature vectors for account nodes and fuses them with generic graph embeddings to learn node representations for fraud account detection.

\item BERT4ETH \cite{bert4eth} pre-trains a Transformer on Ethereum transaction sequences using contrastive masked address prediction to learn transferable account representations for downstream fraud classification.

\item DIAM \cite{DIAM} models multi-chain transaction networks as attributed directed multigraphs and applies dis-crepancy-aware message passing to mitigate anti-homophily for illicit account detection.

\end{itemize}

\begin{figure}[!t]
    \centering
    \includegraphics[width=\linewidth]{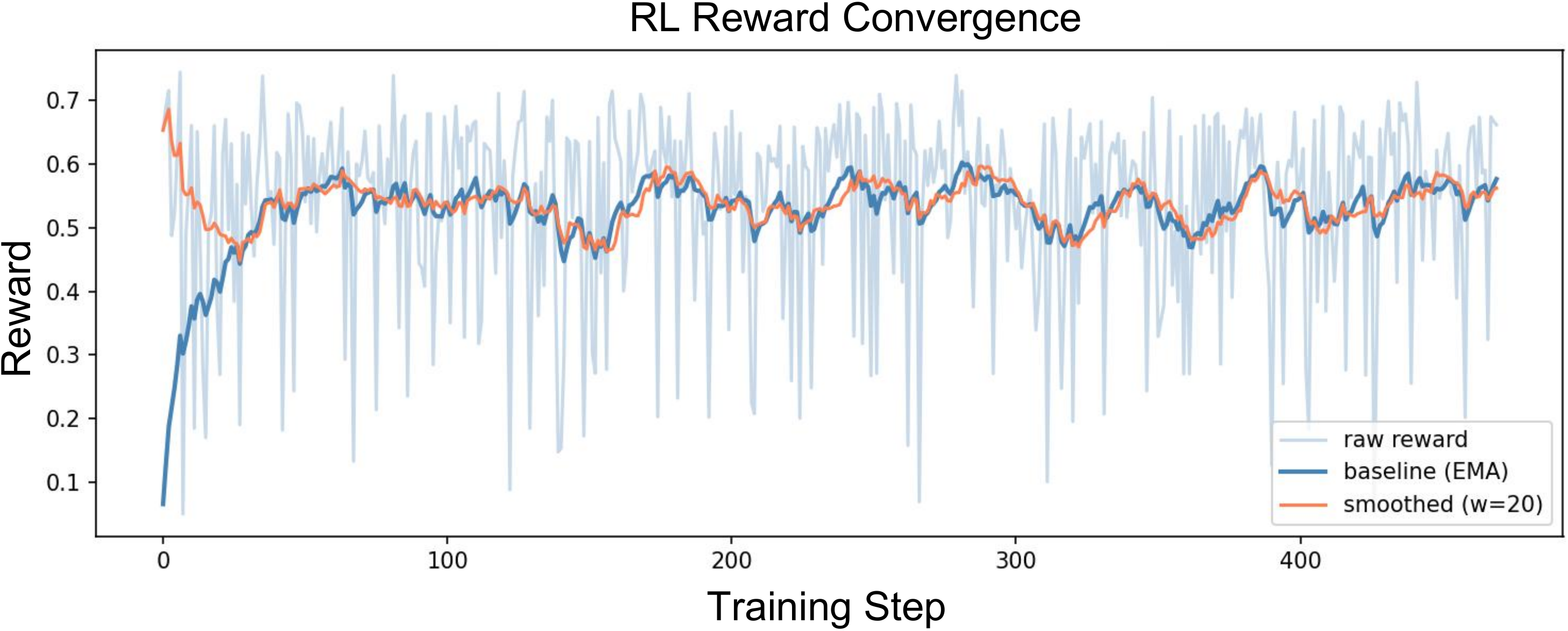}
    \caption{RL reward convergence curve during LLM fine-tuning on the Ethereum dataset. The smoothed reward stabilises rapidly within the first 50 steps, confirming the stability of the proposed RL-based training strategy.}
    \label{fig3}
\end{figure}

\section{Baseline Implementation Settings}
\label{section4}
We use the official implementation code for all open-source baselines. For DeepWalk and Node2Vec, the random walk step size is set to 30, the number of walks for each account node is set to 200, and the node embedding dimension is set to 64. For GCN, we use a two-layer network with a hidden-layer dimension of 64. For GAT, we set the hidden dimension to 256 and used 8 attention heads. For GIN and GraphSAGE, we employ the hidden-layer dimension of 256. 
In the graph anomaly detection method, for FLAG, we use CARE-GNN as the GNN backbone. In the single-chain fraud detection method, for Ethident, we use avgAmount to guide subgraph sampling, and for TEGDetector, we use max pooling in the pooling layer. In the multi-chain fraud detection method, for BERT4ETH, we use the in/out-separation variant. All other parameters not explicitly specified are set to the recommended values from their respective original works to ensure optimal model performance. 
Furthermore, in the Ethereum dataset, except for the multi-chain fraud detection method, which does not require generating optimal account features through feature engineering and can be used directly as the initial node representation, the other methods follow existing work \cite{msy} and select 15-dimensional account features as the initial node features.

\begin{figure}[!t]
    \centering
    \includegraphics[width=0.7\linewidth]{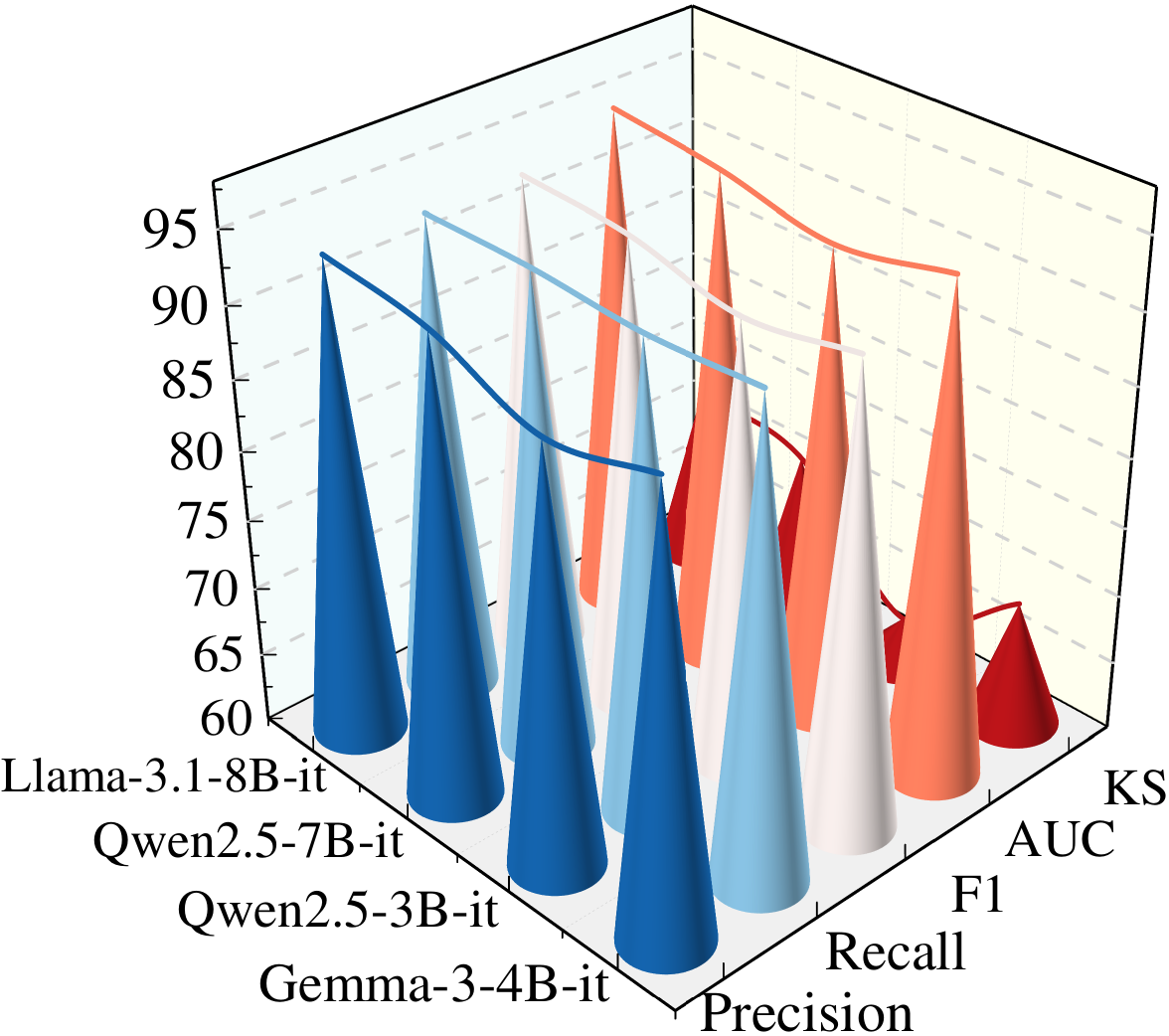}
    \caption{CoT prompt template of the forensics analyst agent.}
    \label{fig5}
\end{figure}

\section{More Discussion}
\label{section5}
$\triangleright$ \textit{Q1. Does the LLM have prior knowledge of account ground-truth labels, such that the observed performance gains may be attributed to label leakage rather than genuine learning?}

To prevent label leakage arising from the LLM's pre-trained knowledge, all account identifiers are removed from the transaction data fed into the Forensics Analyst Agent, and no label information is injected into the CoT prompts of any agent. Furthermore, we explicitly prohibit the LLM from retrieving or referencing any externally accessible information, including publicly disclosed fraud reports or labelled account databases, thereby preventing the model from associating account identifiers with known labels and ensuring the reliability of the detection results.

$\triangleright$ \textit{Q2. How stable is the RL fine-tuning process for the agent's underlying LLM, and is there dedicated analysis of the reward convergence during training?}

Figure \ref{fig3} illustrates the reward convergence curve of the RL fine-tuning process on the Ethereum dataset. The smoothed reward rises rapidly from near zero to approximately 0.5 within the first 50 training steps. Subsequently, it stabilises in the range 0.5–0.6 throughout the remainder of training, with no signs of divergence or collapse. The persistent variance in the raw reward reflects the inherent difficulty heterogeneity across subgraphs rather than training instability, and the EMA baseline effectively stabilises the advantage estimation by providing a dynamic reference point. These results confirm that the proposed RL-based fine-tuning strategy converges reliably and remains stable across the entire training process.

$\triangleright$ \textit{Q3. Which LLM backbone is most suitable for the proposed agents in terms of both detection performance and computational efficiency?}

Figure \ref{fig5} presents the detection performance of UniDetect under different LLM backbones. Overall, Gemma-3-4B-it achieves the most balanced performance across all five metrics. Llama-3.1-8B-it yields slightly higher Precision but at the cost of approximately twice the parameter count and training overhead. Qwen2.5-7B-it achieves comparable $F_1$ and AUC to Gemma-3-4B-it but shows a noticeable drop in KS, suggesting weaker separability between fraudulent and benign accounts. Qwen2.5-3B-it exhibits the lowest performance across most metrics, particularly in KS, indicating that an excessively small backbone limits the quality of generated transaction summaries. These results confirm that Gemma-3-4B-it strikes a favourable balance between detection effectiveness and computational efficiency, justifying its selection as the default backbone in UniDetect.

\end{document}